\documentclass[pre,aps,showpacs,nofootinbib,superscriptaddress]{revtex4}
\usepackage{graphicx}
\usepackage{natbib}
\usepackage{bm}
\usepackage{amssymb}
\usepackage{amsmath}
\usepackage{euscript}
\usepackage{esint}
\usepackage{prelim2e}
\usepackage{slashed}

\begin{document}

\title{Premixed Flame Propagation in Curved Channels}

\author{Hazem El-Rabii}
\author{Guy Joulin}

\affiliation{%
Laboratoire de Combustion et de D\'etonique, CNRS/ENSMA, 1 av.
Cl\'ement Ader, 86961
Futuroscope, Poitiers, France}%

\author{Kirill A. Kazakov}
\affiliation{%
Department of Theoretical Physics, Physics Faculty, Moscow State
University, 119991, Moscow, Russian Federation}

\begin{abstract}
A theory of flame propagation in curved channels is developed within the framework of the on-shell description of premixed flames. Employing the Green function appropriate to the given channel geometry, an implicit integral representation for the burnt gas velocity is constructed. It is then used to derive an explicit expression for rotational component of the gas velocity near the flame front by successive separation of irrotational contributions. We prove that this separation can be performed in a way consistent with boundary conditions at the channel walls. As a result, the unknown irrotational component can be projected out by applying a dispersion relation, thus leading to a closed system of equations for the on-shell fresh gas velocity and the flame front position. These equations show that in addition to the usual nonlocality associated with potential flows, vorticity produced by a curved flame leads to specific nonlocal spatial and temporal influence of the channel geometry on the flame evolution. To elucidate this influence, three special cases are considered in more detail -- steady flame stabilized by incoming flow in a bottle-shaped channel, quasi-steady flame, and unsteady flame with small gas expansion propagating in a channel with slowly varying width. In the latter case, analytical solutions of the derived equations are obtained in the first post-Sivashinsky approximation using the method of pole decomposition.
\end{abstract}
 \maketitle

\section{Introduction}

Analytical description of premixed flame propagation is one of the most difficult and exiting tasks in combustion science. In its full generality, it represents a tremendously complicated problem involving several length scales, nonlocal interactions, and various instabilities inherent in the governing mechanisms, which in turn necessitate consideration of flame dynamics in the regime of saturated nonlinearity. On the other hand, this problem naturally admits several important simplifications which make it amenable to theoretical treatment. First of all, the fact that the deflagration is an essentially subsonic process allows the gas flows be considered incompressible, both in the fresh and burnt gas regions. Also, neglecting viscous effects in these regions, gases can be treated as ideal outside the flame front. These assumptions turn out to be very good approximations to real flames.  Furthermore, large separation of scales determining inner dynamics of the flame front and its global behavior allows the use of asymptotic methods to derive local relations between the fresh and burnt gas flows at the flame front. These relations contain information about transport processes inside the front, and are expressed in the form of jump conditions for the gas velocity and its pressure across the front, and an equation determining local consumption rate (the evolution equation). They play the role of boundary conditions for the large-scale flow at the front considered as a surface of discontinuity.

Yet, even simplified this way the problem is still exceedingly complicated. It requires solving a system of nonlinear partial differential equations on both sides of the flame front, to be chosen so as to satisfy the jump conditions
across the moving front. It is clearly impossible to resolve this system explicitly, because even if the flame propagates in an initially quiescent fluid, so that the gas flow is potential upstream, this is no longer true downstream the flame because of vorticity generated by the curved flame front. In this connection, it is important to note that flames used in practice are characterized by large thermal expansion of gases: the fresh to burnt gas density ratio, $\theta,$ is normally $5\div 10.$ It is not difficult to show \cite{kazakov3} that under such circumstances, the weak nonlinearity approximation is not applicable to the description of flame evolution (except the earlier stage of development of the Darrieus-Landau instability \cite{darrieus,landau}), so that the problem of solving the flow equations is faced in its full generality.

This difficulty precludes theoretical analysis if one is concerned with the explicit structure of the burnt gas flow. However, it is evolution of the flame front, its position and shape, which is of greatest interest in practice. As was shown in \cite{kazakov1,kazakov2,jerk1,jerk2}, this limitation of the problem opens a way round the above-mentioned difficulty. Namely, it turns out to be possible to derive an exact system of equations describing the flame-front evolution, which is closed in the sense that it involves only restrictions of the flow variables to the front (their {\it on-shell} values). Not only this approach permits the well-known results on the flame-front dynamics be derived anew in a simple and elegant way, it also brings to the scope of theoretical treatment classical problems which have not been accessible by means of the conventional analysis \cite{jerk3,kazakov4}.

Considerations of \cite{kazakov1,kazakov2,jerk1,jerk2,jerk3,kazakov4} have been restricted to the simplest case of flames propagating in straight channels. A question of principle which is very important from both experimental and theoretical points of view is whether the results obtained can be extended to channels with curved walls. Importance of this question lies in the fact already mentioned that the flame evolution is an essentially nonlocal process. In view of this nonlocality, flame propagation is naturally expected to be affected by the wall curvature. This effect is anticipated to be especially pronounced in the two-dimensional case due to peculiar long-range behavior of the Green functions.

The purpose of this paper is to show that in the two-dimensional case, the above question resolves in the positive. Namely, the on-shell flame description will be extended to symmetric channels with arbitrarily curved walls. The influence of channel geometry on flame dynamics turns out to be much more complicated than predicted by the potential flow models \cite{siv1,frankel1990}. Because of vorticity produced by the curved flame, this influence does not reduce to a mere conformal transformation of the flow variables. Apart from this fact which is clear from the outset, our investigation reveals specific spatial and temporal nonlocalities associated with the variable channel shape.

The paper is organized as follows. Derivation of the main integro-differential equation for the on-shell gas velocity, called for brevity the master equation, proceeds along the lines of Refs.~\cite{kazakov1,jerk1}. We first derive in Sec.~\ref{integrep} an implicit integral representation for rotational component of the burnt gas velocity using the Green function approach, and discuss its validity for various asymptotic channel geometries. In Sec.~\ref{dispersrel}, we generalize the dispersion relation for irrotational velocity component to arbitrarily curved channels. This consideration identifies conditions under which irrotational velocity can be safely neglected, which is then used in Sec.~\ref{generalst} to gradually simplify the integral formula for the rotational velocity component. An explicit expression for this component at the flame front is obtained in Sec.~\ref{nearfront}, which is the last step of the derivation of the master equation written down in Sec.~\ref{mastereq}. To elucidate the structure of this equation, we consider three special cases -- steady flame stabilized by incoming flow in a bottle-shaped channel (Sec.~\ref{steady}), a quasi-steady flame, and a flame with $(\theta - 1)\ll 1$ propagating in a channel with slowly varying width (Sec.~\ref{smalltheta}), which admits complete analytical investigation. The results of the work are discussed in Sec.~\ref{conclude}.

\section{Integral representation of the flow velocity}\label{integrep}

\begin{figure}
\centering
\includegraphics[width=0.8\textwidth]{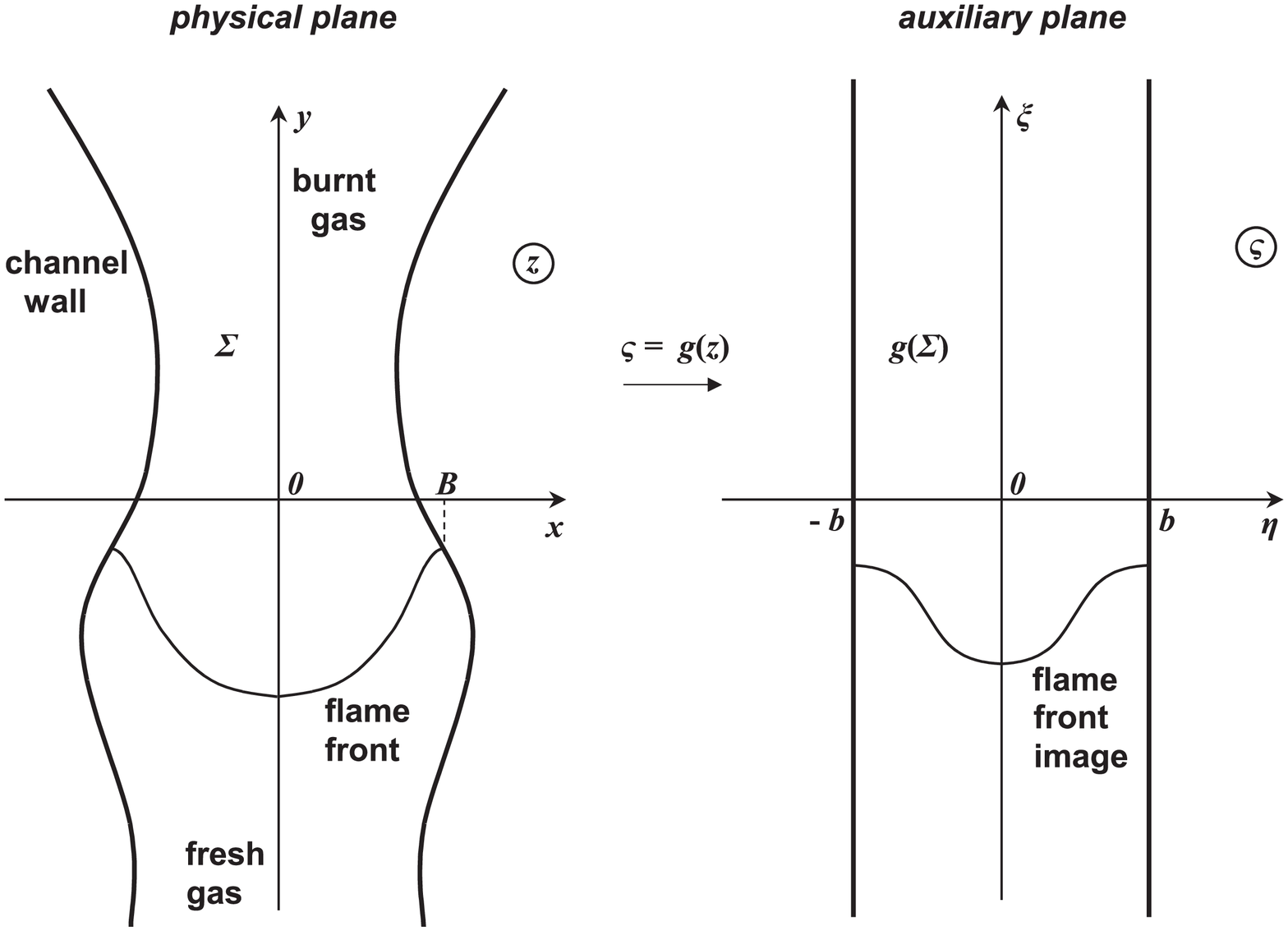}
\caption{}\label{fig1}
\end{figure}

Consider a $2D$ flame propagating in a symmetric channel filled with an initially quiescent uniform ideal reactive fluid. The channel width is assumed to vary smoothly along the channel, but is otherwise arbitrary. Let the Cartesian coordinates $(x,y)$ be chosen so that the $y$-axis is along the symmetry axis, $y = - \infty$ being in the fresh gas. Then the flame-front position at time $t$ can be described by an equation $y=f(x,t).$ The fluid velocity $(w,u)$ will be measured in units of the planar flame speed relative to the fuel, and the fluid density in units on the fuel density, $\theta>1$ denoting its ratio to that of the burnt gas. In the subsequent analysis, we shall widely use complex notations combining the coordinates and  velocity components into the complex variables $z = x + iy$ and $\omega = u + iw.$ In addition to the physical $z$-plane, it will be convenient to introduce an auxiliary $\zeta$-plane, $\zeta = \eta + i\xi,$ with $\eta,\xi$ real, such that the function $\zeta = g(z)$ maps the curved channel in the physical plane onto a straight channel of constant width, $2b,$ in the auxiliary plane. In coordinate form, this mapping will be written simply as $\eta = \eta(x,y),$ $\xi = \xi(x,y),$ and similarly for its inverse. We assume that $g(z)$ is analytic on some open subset in $z$-plane, containing the physical channel, and that $g(iy) = i\xi,$ $\xi = -\infty$ corresponding to the fresh gas (see Fig.~\ref{fig1}). The symmetry of the physical channel with respect to the $y$-axis implies that the function $g(z)$ possesses the following reflection property
\begin{eqnarray}\label{reflect}
g(-\bar{z}) = - \overline{g(z)}\,,
\end{eqnarray}
\noindent where the bar denotes complex conjugation. Accordingly, we will be concerned with symmetrical flame configurations satisfying
\begin{eqnarray}\label{reflect1}
f(x,t) = f(-x,t),\quad w(x,y,t) =
- w(-x,y,t), \quad u(x,y,t) = u(-x,y,t).
\end{eqnarray}
\noindent
Under the assumptions that the gas flow is ideal incompressible, its velocity obeys the following equations in the bulk
\begin{eqnarray}\label{flow1}
\frac{\partial w}{\partial x} + \frac{\partial u}{\partial y} &=& 0\,,\\
\frac{\partial\sigma}{\partial t} + w \frac{\partial\sigma}{\partial x} + u \frac{\partial\sigma}{\partial y} &=& 0 \,,\label{flow2}
\end{eqnarray}\noindent where
\begin{eqnarray}
\sigma &=& \frac{\partial u}{\partial x} - \frac{\partial w}{\partial y}\label{flow3}
\end{eqnarray}\noindent is the vorticity. Let us introduce the stream function $\psi$ according to
\begin{eqnarray}\label{psi}
w = - \frac{\partial\psi}{\partial y}\,, \quad u = \frac{\partial\psi}{\partial x}\,.
\end{eqnarray}\noindent
Then the continuity equation (\ref{flow1}) becomes an identity, while the definition (\ref{flow3}) takes the form
\begin{eqnarray}\label{lappsi}
\triangle\psi = \sigma\,, \quad \triangle \equiv \frac{\partial^2}{\partial x^2} + \frac{\partial^2}{\partial y^2}\,.
\end{eqnarray}\noindent
In order to construct an implicit integral representation for the burnt gas velocity, let us treat the latter relation as the Poisson equation for $\psi,$ considering the vorticity distribution downstream of the flame front as a given function. Since the stream function must be constant along the channel walls, we have to find the Green function, $G(x,y;\tilde{x},\tilde{y}),$ of the Laplace operator, satisfying
$$\triangle G(x,y;\tilde{x},\tilde{y}) = \delta(x-\tilde{x})\delta(y-\tilde{y})\,, \quad G(x(\pm b,\xi),y(\pm b,\xi);\tilde{x},\tilde{y}) = {\rm const}\,,$$ where $\delta(\cdot)$ is the Dirac function. It is not difficult to verify that one can choose
\begin{eqnarray}\label{greenf}
G(x,y;\tilde{x},\tilde{y}) = \frac{1}{4\pi}\ln\frac{\displaystyle\sin\left\{\frac{\pi}{4b}\left[g(z) - g(\tilde{z})\right]\right\}}{\displaystyle\cos\left\{\frac{\pi}{4b}\left[g(z) + \overline{g(\tilde{z})}\right]\right\}} + {\rm c.c.}\,,
\end{eqnarray}\noindent where ``c.c.'' stands for the complex conjugate of the preceding expression. Indeed, the right hand side of Eq.~(\ref{greenf}) is analytic everywhere in the channel except the point $z=\tilde{z},$ therefore, it satisfies the Laplace equation. On the other hand, in a vicinity of $z=\tilde{z}$
$$G(x,y;\tilde{x},\tilde{y}) = \frac{1}{4\pi}\ln(z - \tilde{z}) + R(z,\tilde{z}) + {\rm c.c.},$$ where $R(z,z')$ is analytic at $z=z'.$ Hence, denoting $r = \sqrt{(x-\tilde{x})^2+(y-\tilde{y})^2} = \sqrt{(z - \tilde{z})\overline{(z - \tilde{z})}},$ one has $G(x,y;\tilde{x},\tilde{y}) = \frac{1}{2\pi}\ln r$ plus terms satisfying the Laplace equation, so that $G$ correctly reproduces the $\delta$-singularity at $(x,y) = (\tilde{x},\tilde{y}).$ Finally, it is not difficult to check that $G(x(\pm b,\xi),y(\pm b,\xi);\tilde{x},\tilde{y}) = 0.$
Thus, the general solution of Eq.~(\ref{lappsi}) can be written as
\begin{eqnarray}\label{integral}
\psi(x,y) = \int_{\Sigma}dsG(x,y;\tilde{x},\tilde{y})\sigma(\tilde{x},\tilde{y}) + \psi_0(x,y),
\end{eqnarray}\noindent
where $\Sigma$ denotes the (open) region downstream the flame front, $ds = d\tilde{x}d\tilde{y},$ and $\psi_0(x,y)$ the general solution of $\triangle \psi = 0,$ satisfying the boundary conditions
\begin{eqnarray}\label{boundcond}
\psi_0(x(- b,\xi),y(- b,\xi)) = {\rm const}_-\,, \quad \psi_0(x(+ b,\xi),y(+ b,\xi)) = {\rm const}_+\,.
\end{eqnarray}\noindent The difference of the constants ${\rm const}_{\pm}$ gives  the overall mass flux of the burnt gas. Following the usual procedure \cite{kazakov2,jerk2}, we now rewrite the first term in
Eq.~(\ref{integral}) as an integral over fluid particle trajectories. Namely, at
each given time instant $t,$ we consider $\Sigma$ as spanned by fluid elements that crossed the flame front during its evolution up to the given $t.$ Symbolizing differentiation with respect to $x$ by a prime, let us introduce the quantities $N = (1 + f'^2)^{1/2}$ and $\hat{v}^n_+ = (\hat{\bm{v}}_+,\bm{n}),$
where $$\hat{\bm{v}}_+ =
(w_+,\hat{u}_+)\,, \quad \hat{u}_+(x,t) \equiv u_+(x,t) -
\frac{\partial f(x,t)}{\partial t}\,,$$ $\bm{n}$ is the unit vector normal to the flame front (pointing towards the burnt gas). The subscript ``$+$'' (``$-$'') is used throughout to indicate restriction to the front of the corresponding downstream (upstream) flow variable; thus, $\hat{v}^n_+$ is the normal burnt gas velocity relative to the flame front. Then, taking into account that the ``volume'' $ds$ of
an element is conserved because of flow incompressibility, as is the vorticity [Cf. Eq.~(\ref{flow2})], we can write
\begin{eqnarray}\label{vint1}&&
\int_{\Sigma}ds G(x,y;\tilde{x},\tilde{y})\sigma(\tilde{x},\tilde{y}) \equiv \psi^v(x,y)\\&& = \int_{-B}^{+B}d\tilde{x} \int_{-\infty}^{t}d\tau
N(\tilde{x},\tau)\hat{v}^n_+(\tilde{x},\tau)
\sigma_+(\tilde{x},\tau)\frac{1}{4\pi}\ln\frac{\displaystyle\sin\left\{\frac{\pi}{4b}\left[g(z) - g(Z(\tilde{x},t,\tau))\right]\right\}}{\displaystyle\cos\left\{\frac{\pi}{4b}\left[g(z) + \overline{g(Z(\tilde{x},t,\tau))}\right]\right\}} + {\rm c.c.} \,,
\nonumber
\end{eqnarray}
\noindent where $B=B(t)$ is the abscissa of the right end-point of the flame front (it is the solution of $\eta(B + if(B,t)) = b$), and $Z(\tilde{x},t,\tau) = X(\tilde{x},t,\tau) + iY(\tilde{x},t,\tau)$ the current
position (in $z$-plane) of a fluid element that crossed the point $(\tilde{x},
f(\tilde{x},\tau))$ at the flame front at time $\tau\le t.$ Using Eq.~(\ref{reflect}), and taking into account that $\sigma_+, X$ are odd functions of $\tilde{x},$ while $N,\hat{v}^n_+,$ and $Y$ are even, it is not difficult to see that $\psi^v(-x,y) = - \psi^v(x,y),$ so that the corresponding velocity field satisfies Eq.~(\ref{reflect1}), and therefore, so does the irrotational contribution generated by $\psi_0.$

To ensure convergence of the integral over $\tau,$ one has to restrict the class of  allowed functions $g(z),$ and/or introduce some assumption about the flow at infinity. The simplest choice is to assume that the channel is asymptotically straight, {\it i.e.,} $g(z) \to {\rm const}\cdot z$ for ${\rm Im}\,z \to + \infty,$ with an initially planar flame propagating from infinity. Then $\sigma_+(\tilde{x},\tau)$ is exponentially small for $\tau \to -\infty.$ However, this assumption is very strong and can be relaxed. In fact, it may be sufficient to have an asymptotically V-shaped  channel, with the opening angle $2\alpha < \pi.$ Indeed, in this case $$g(z) = \frac{ib}{\alpha}\ln\frac{z}{i}\,, \quad {\rm Im}\,z \to +\infty,$$ so that for large $|\tau|,$ the logarithm in the integrand of Eq.~(\ref{vint1}) is proportional to
$$\exp\left\{\frac{\pi}{4b}\left[-\frac{2b}{\alpha}\ln(u_{\infty}|\tau|) - 2ig(z)\right]\right\} \sim (u_{\infty}|\tau|)^{-\pi/2\alpha}\,,$$ where $u_{\infty} = \partial Y(\tilde{x},t,\tau)/\partial t.$ Thus, if $\alpha<\pi/2,$ the $\tau$-integral converges (since the other factors in the integrand are bounded functions of $\tau$), {\it provided that $u_{\infty}$ does not vanish for the gas elements carrying nonzero vorticity.} But the latter condition is just what is normally observed in experiments with Bunsen flames, which indicate that vorticity produced by the flame is carried away in thin jets spanned by the stream lines with $u_{\infty}\ne 0.$ Restrictions as to the asymptotic channel geometry can be weaken even further taking into account parity properties of the flow variables.

The last comment concerns analytical properties of the flow variables. As always, we assume that the flame is stable with respect to short wavelength disturbances {\it i.e.,} that there is a short wavelength cutoff, $\lambda_{\rm c},$ whose precise value is determined by specific transport processes inside the flame front. This cutoff ensures smoothness of the functions involved, which are considered throughout as infinitely differentiable.

\section{Derivation of the master equation}

\subsection{Dispersion relation for irrotational velocity component}\label{dispersrel}

It is impossible to compute the right hand side of Eq.~(\ref{vint1}) in general, because for that one would had to solve the system of hydrodynamic equations downstream explicitly. However, in order to describe dynamics of the flame it is sufficient to determine the flow structure in a vicinity of the front only. This can be done by decomposing the burnt gas velocity as $$\bm{v} = \bm{v}^p + \bm{v}^v\,,$$ where $\bm{v}^p$ is a potential component incorporating the unknown information about the bulk flow, which is eventually ``projected out'' using a dispersion relation. To derive this relation, we require $\bm{v}^p$ to satisfy the following conditions:
\begin{itemize}
\item[a)]{${\rm div}\bm{v}^p = 0,$ ${\rm rot}\bm{v}^p = 0,$ implying that $\omega^p = u^p + iw^p$ is analytic function of $z\in \Sigma,$}
\item[b)]{the function $\Omega(\zeta) = \omega^p(g^{-1}(\zeta))\frac{\displaystyle dg^{-1}(\zeta)}{\displaystyle d\zeta}$ is bounded in $g(\Sigma),$}
\item[c)]{$\Omega(-b+i\xi) = \Omega(+b+i\xi)$ for all $\xi$ downstream.}
\end{itemize}
Consider the region downstream of the flame front in the auxiliary plane. Since $\omega^p(z)$ is assumed analytic downstream, $\Omega(\zeta)$ is analytic for $\zeta \in g(\Sigma),$ so that one has, by virtue of the Cauchy theorem,
\begin{eqnarray}\label{cauchy}
\frac{1}{4b i}\int_{\partial g(\Sigma)}d\tilde{\zeta} \cot\left\{\frac{\pi}{2b}(\tilde{\zeta} - \zeta)\right\}\Omega(\tilde{\zeta}) = \Omega(\zeta)\,,
\end{eqnarray}
\noindent
where $\partial g(\Sigma)$ is the boundary of $g(\Sigma),$ ran counterclockwise by $\tilde{\zeta}.$ If $\Omega(\zeta)$ satisfies the condition c),
the wall contributions to the contour integral cancel each other because of the $2b$-periodicity of the cotangent. If, in addition to that, b) is also met, then taking the limit $\zeta \to \zeta_+ = g(z_+),$ $z_+ = x + if(x,t),$ in Eq.~(\ref{cauchy}) gives
$$\frac{1}{2b i}\fint_{g(F)}d\tilde{\zeta} \cot\left\{\frac{\pi}{2b}(\tilde{\zeta} - \zeta_+)\right\}\Omega(\tilde{\zeta}) - \frac{1}{2bi}\int_{-b+i\infty}^{+b+i\infty}d\tilde{\zeta}\,\Omega(\tilde{\zeta}) = \Omega_+\,,$$ where the integral over the image $g(F)$ of the flame front is understood in the principal value sense. We can get rid off the constant contribution  of the infinitely remote interval by differentiating this equation with respect to $x.$ After that, integration of the first term by parts taking into account condition c) yields
\begin{eqnarray}\label{dispers}
\frac{g'(z_+)}{2b i}\fint_{g(F)}d\tilde{\zeta} \cot\left\{\frac{\pi}{2b}(\tilde{\zeta} - \zeta_+)\right\}\frac{d\Omega(\tilde{\zeta})}{d\tilde{\zeta}} = \left(\Omega_+\right)'\,, \quad g'(z_+) \equiv \frac{d}{dx}g(x + if(x,t))\,,
\end{eqnarray}
\noindent or equivalently,
\begin{eqnarray}\label{dispers1}
\frac{g'(z_+)}{2b i}\fint_{-B}^{+B}d\tilde{x} \cot\left\{\frac{\pi}{2b}[g(\tilde{z}) - g(z_+)]\right\}(\tilde{\Omega}_+)' = \left(\Omega_+\right)'\,,
\end{eqnarray}
\noindent where $\tilde{z} = \tilde{x} + if(\tilde{x},t)\,, \tilde{\Omega}_+ \equiv \Omega(g(\tilde{z})).$

\subsection{General structure of rotational velocity component}\label{generalst}

We now proceed to the derivation of an explicit expression for the rotational velocity component, $\bm{v}^v.$ Let the equality of two functions $\varphi_1(x,y),\, \varphi_2(x,y)$ up to a field satisfying a) -- c) be symbolized as $\varphi_1\stackrel{\circ}{=} \varphi_2.$ Then Eqs.~(\ref{integral}), (\ref{vint1}) give for the complex burnt gas velocity
\begin{eqnarray}\label{vint2}&&
\omega \stackrel{\circ}{=} \frac{\slashed{\partial}}{4\pi}\left[\int_{-B}^{+B}d\tilde{x} \int_{0}^{+\infty}d\tau
M(\tilde{x},t-\tau)\ln\frac{\displaystyle\sin\left\{\frac{\pi}{4b}\left[g(z) - g(Z(\tilde{x},t,t-\tau))\right]\right\}}{\displaystyle\cos\left\{\frac{\pi}{4b}\left[g(z) + \overline{g(Z(\tilde{x},t,t-\tau))}\right]\right\}} + {\rm c.c.}\right],
\nonumber\\
\end{eqnarray}
\noindent where we denoted $\slashed{\partial} \equiv \partial/\partial x - i\partial/\partial y,$ introduced the memory kernel $M(\tilde{x},t) \equiv N(\tilde{x},t)\hat{v}^n_+(\tilde{x},t) \sigma_+(\tilde{x},t),$ and changed the integration variable $\tau \to t - \tau.$ In fact, condition a) is obviously fulfilled by the velocity $\bm{v}_0$ corresponding to the term $\psi_0$ in Eq.~(\ref{integral}). Furthermore, under the assumption about convergence of the $\tau$-integral, discussed at the end of Sec.~\ref{integrep}, the right hand side of Eq.~(\ref{vint2}) is bounded in $\Sigma,$ and therefore, so is the velocity $\omega_0 = u_0 + iw_0,$ because the total burnt gas velocity is bounded. Hence, the function $\Omega_0(\zeta) = \omega_0(g^{-1}(\zeta))\frac{\displaystyle dg^{-1}(\zeta)}{\displaystyle d\zeta}$ is bounded in $g(\Sigma),$ so that condition b) is met. Finally, to show that c) is also fulfilled, we write $$\frac{dg^{-1}(\zeta)}{\displaystyle d\zeta} = -i\frac{\partial x}{\partial\xi} + \frac{\partial y}{\partial\xi}\,,$$ and then
$$\Omega_0 = u_0\frac{\partial y}{\partial\xi} + w_0\frac{\partial x}{\partial\xi}+
i\left(w_0\frac{\partial y}{\partial\xi} - u_0\frac{\partial x}{\partial\xi}\right)\,.$$ Differentiating the boundary condition (\ref{boundcond}) with respect to $\xi,$ one finds
$$\left(w_0\frac{\partial y}{\partial\xi} - u_0\frac{\partial x}{\partial\xi}\right)(\pm b,\xi) = 0\,,$$ which shows that the imaginary part of $\Omega_0$ vanishes at the walls. On the other hand, differentiation of the reflection property (\ref{reflect}) written in the form $x(-\eta,\xi) = - x(\eta,\xi),$ $y(-\eta,\xi) = y(\eta,\xi),$ gives $$\frac{\partial x}{\partial\xi}(-b,\xi) = - \frac{\partial x}{\partial\xi}(+b,\xi)\,, \quad \frac{\partial y}{\partial\xi}(-b,\xi) = \frac{\partial y}{\partial\xi}(+b,\xi)\,.$$ Together with $u_0(-x,y) = u_0(x,y),$ $w_0(-x,y) =  - w_0(x,y),$ these relations prove that $\Omega_0(-b+i\xi) = \Omega_0(+b+i\xi).$

Next, consider the kernel of the $\tilde{x}$-integral in Eq.~(\ref{vint2})
\begin{eqnarray}&&\label{kernel}
K(\tilde{x},t,\zeta) \equiv \frac{1}{2\pi}\int_{0}^{+\infty}d\tau
M(\tilde{x},t-\tau)\ln\frac{\displaystyle\sin\left\{\frac{\pi}{4b}\left[\zeta - g(Z(\tilde{x},t,t-\tau))\right]\right\}}{\displaystyle\cos\left\{\frac{\pi}{4b}\left[\zeta + \overline{g(Z(\tilde{x},t,t-\tau))}\right]\right\}} + {\rm c.c.}
\end{eqnarray}\noindent
Our task is to evaluate this function in the case when the observation point $z$ is close to the flame front. To that end, we have to perform first some general transformations on $K,$ which is the subject of the present section. Before we proceed to the computation, let us prove the following simple but important \\ {\it Lemma}: if the complex-valued function $p=p(\tilde{x},\mu)$  ($\mu$ is a parameter) satisfying $p(-\tilde{x},\mu) = - \overline{p(\tilde{x},\mu)},$ is such that the point $g(p)\notin g(\Sigma),$ and if  $m(\tilde{x})$ is odd, then the field
\begin{eqnarray}&&\label{lemma1}
\omega^p = \slashed{\partial}\left[\int_{-B}^{+B}d\tilde{x}
m(\tilde{x})\ln\frac{\displaystyle\sin\left\{\frac{\pi}{4b}\left[g(z) - g(p)\right]\right\}}{\displaystyle\cos\left\{\frac{\pi}{4b}\left[g(z) + \overline{g(p)}\right]\right\}} + {\rm c.c.}\right]
\end{eqnarray}\noindent satisfies conditions a)--c). Indeed, noting that for $\varphi(z)$  analytic,
\begin{eqnarray}&&\label{diff}
\slashed{\partial}\varphi(z) = 2\frac{d \varphi(z)}{dz}\,, \quad \slashed{\partial}\,\overline{\varphi(z)} = 0\,,
\end{eqnarray}\noindent
we see that $\omega^p(z)$ is analytic in $\Sigma.$ Furthermore, the corresponding function $\Omega^p = \omega^p(g^{-1}(\zeta))dg^{-1}/d\zeta$ takes the form
$$\Omega^p(\zeta,z_1,\tau) = 2\frac{d}{d\zeta}\int_{-B}^{+B}d\tilde{x}
m(\tilde{x})\ln\frac{\displaystyle\sin\left\{\frac{\pi}{4b}\left[\zeta - g(p)\right]\right\}}{\displaystyle\cos\left\{\frac{\pi}{4b}\left[\zeta + \overline{g(p)}\right]\right\}}.$$ It is bounded for $\zeta \in g(\Sigma),$ so that b) is also met. Finally, c) is easily verified changing the integration variable $\tilde{x} \to - \tilde{x},$ and taking into account oddness of the function $m(\tilde{x}).$ As a consequence of this result, differentiation of Eq.~(\ref{lemma1}) with respect to $\mu$ shows that the function
\begin{eqnarray}&&\label{lemma2}
\slashed{\partial}\int_{-B}^{+B}d\tilde{x}
m(\tilde{x})\left(\cot\left\{\frac{\pi}{4b}\left[\zeta - g(p)\right]\right\}\frac{\partial g(p)}{\partial \mu} - \tan\left\{\frac{\pi}{4b}\left[\zeta + \overline{g(p)}\right]\right\}\overline{\frac{\partial g(p)}{\partial \mu}}\right) \nonumber
\end{eqnarray}\noindent also satisfies a)--c).

We now rewrite the kernel integrating it by parts. For this purpose, we introduce the quantity
\begin{eqnarray}\label{eumdef}
\int\limits_{\tau}^{+\infty}d\tau_1 M(\tilde{x},t-\tau_1) \equiv
\EuScript{M}(\tilde{x},t,\tau)\,,
\end{eqnarray}
\noindent and abbreviate $Z(\tilde{x},t,t-\tau) \equiv Z_{\tau},$ $g(Z_{\tau})\equiv g_{\tau}.$ This yields
\begin{eqnarray}\label{vint4}&&
K(\tilde{x},t,\zeta) = \EuScript{M}(\tilde{x},t,0)\frac{1}{2\pi}\ln\frac{\displaystyle\sin\left\{\frac{\pi}{4b}\left[\zeta - g(\tilde{z})\right]\right\}}{\displaystyle\cos\left\{\frac{\pi}{4b}\left[\zeta + \overline{g(\tilde{z})}\right]\right\}} \nonumber\\&& - \frac{1}{8b}\int_{0}^{+\infty}d\tau
\EuScript{M}(\tilde{x},t,\tau)\left(\cot\left\{\frac{\pi}{4b}\left[\zeta - g_{\tau}\right]\right\}\dot{g}_{\tau} - \tan\left\{\frac{\pi}{4b}\left[\zeta + \overline{g}_{\tau}\right]\right\}\dot{\overline{g}}_{\tau}\right) + {\rm c.c.},
\end{eqnarray}
\noindent where the dot denotes differentiation with respect to $\tau.$
By virtue of the Lemma, the first term here can be omitted,\footnote{This will be symbolized by the same sign $\stackrel{\circ}{=}$ used previously to relate velocity fields.} because $\Sigma$ is open, while  $\tilde{z}\in F\slashed{\subset}\Sigma.$

\begin{figure}
\centering
\includegraphics[width=0.6\textwidth]{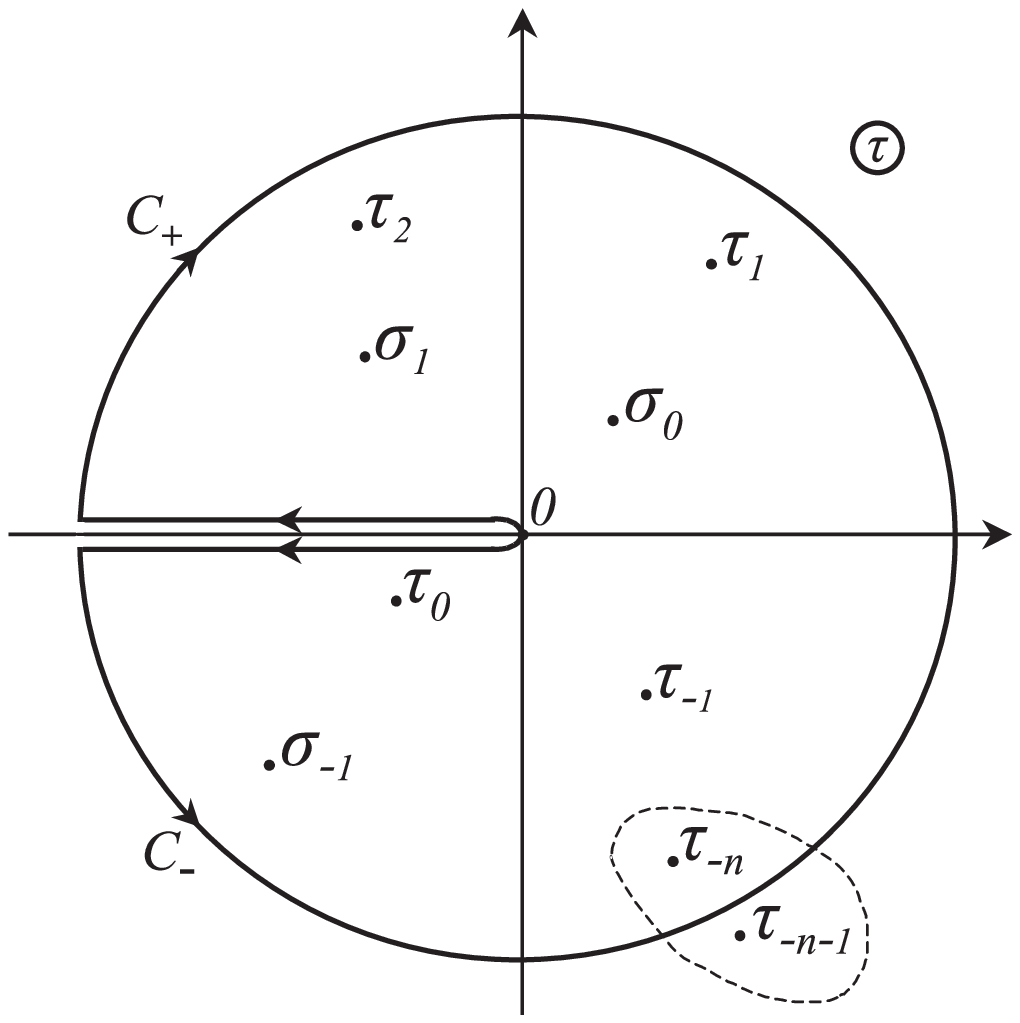}
\caption{}\label{fig2}
\end{figure}

Let us allow $\tau$ to take complex values. Under the assumption of existence of a short wavelength cutoff, all flow variables are smooth functions of time. Hence, these functions (in particular, the memory kernel $M$) are analytic in a vicinity of the real axis in the complex time-plane. Therefore, the only singularity of the integrand in Eq.~(\ref{vint4}), which can cross the real axis, is at $\tau$ such that
\begin{eqnarray}\label{point}
g(Z_{\tau}) = \zeta.
\end{eqnarray}
\noindent
Specifically, the imaginary part of $\tau$ vanishes at the point $\{\tau,\tilde{x}\}$ satisfying
\begin{eqnarray}\label{point1}
X[\tilde{x},t,t-\tau] = x\,, \quad Y[\tilde{x},t,t-\tau] = y\,.
\end{eqnarray}
\noindent The pole corresponding to $\tau$ satisfying this equation determines the  structure of the rotational component of the burnt gas velocity. We extract the pole contribution by deforming the contour of integration over $\tau.$ Specifically, we write $\int d\tau = 1/2\int d\tau+ 1/2\int d\tau,$ and deform the contour of integration in the two halves symmetrically with respect to the real axis, moving it pass the solution of Eq.~(\ref{point1}), for all $\tilde{x}$ (see Fig.~\ref{fig2}, where $C_+$ and $C_-$ denote the upper and lower branches of the contour, respectively). Since the integrand in Eq.~(\ref{vint4}) involves $\overline{g}_{\tau},$ in order to promote $\tau$ to a complex variable the operation of complex conjugation is to be further specified. It is convenient to define $\overline{g}_{\tau}$ as an analytic continuation of $\overline{(g_{\overline{\tau}})}$ with respect to $\tau$ (in other words, the argument $\tau$ in $\overline{g}_{\tau}$ is not subjected to the conjugation). The pole contribution to the integral kernel then reads, by virtue of the Cauchy theorem,
$$K_0(\tilde{x},t,\zeta) = \frac{i}{2}\chi\left({\rm Im\,\tau_0} \right)\EuScript{M}(\tilde{x},t,\tau_0)\,,$$ where $\chi(x)$ is the sign function,
$$\chi(x) = \left\{
\begin{array}{cc}
+1,& x>0\,,\\
\phantom{+}0, & x = 0\,,\\
-1,&  x<0\,,
\end{array}
\right.$$ and $\tau_0=\tau_0(\tilde{x},t,\zeta)$ denotes the solution of Eq.~(\ref{point}). To keep the boundary conditions satisfied (see condition c)), we must perform the same contour deformation for both terms in the integrand in Eq.~(\ref{vint4}). But when the observation point is at the channel wall (say, the right), one has $g_{\tau_0} = + b + i\xi,$ hence $\zeta + \overline{g}_{\overline{\tau_0}} = 2b,$ which is a pole of the tangent. In order to guarantee continuity of the extracted contribution, we must proceed with contour deformation so as to embrace the pole of the tangent, again for all $\tilde{x}$ and $\zeta \in \Sigma.$ Let us denote this pole $\sigma_0 = \sigma_0(\tilde{x},t,\zeta).$ Thus,
\begin{eqnarray}\label{reflect2}
\sigma_0(\tilde{x},t,b+i\xi) = \overline{\tau_0(\tilde{x},t,b+i\xi)}.
\end{eqnarray}
\noindent
Furthermore, the change $\tilde{x}\to -\tilde{x}$ transforms $Z_{\tau}\to - \overline{Z}_{\tau}$ ($\tau$ is not conjugated), hence, using the property (\ref{reflect}), $g_{\tau} \to - \overline{g}_{\tau}\,,$ which together with Eq.~(\ref{point}) imply
\begin{eqnarray}
\tau_0(-\tilde{x},t,-b+i\xi) = \overline{\tau_0(\tilde{x},t,b+i\xi)}.\nonumber
\end{eqnarray}
\noindent Combining this with Eq.~(\ref{reflect2}) gives
\begin{eqnarray}\label{reflect3}
\tau_0(-\tilde{x},t,-b+i\xi) = \sigma_0(\tilde{x},t,b+i\xi),
\end{eqnarray}
\noindent In other words, inversion $\tilde{x} \to - \tilde{x}$ maps the $\tau_0$-poles at the left wall to the $\sigma_0$-poles at the right wall. Let us denote $\sigma_{-1} = \sigma_{-1}(\tilde{x},t,\zeta)$ the pole of the tangent which is in the same correspondence with the right-wall value of $\tau_0,$ {\it viz.,}
\begin{eqnarray}
\tau_0(-\tilde{x},t,b+i\xi) = \sigma_{-1}(\tilde{x},t,-b+i\xi).
\end{eqnarray}
\noindent Again, continuity requires the contour to encompass the $\sigma_{-1}$-pole. In turn, this pole merges at the right wall a pole of the cotangent, to be denoted $\tau_{-1}(\tilde{x},t,\zeta),$ with the correspondence given by the inversion
\begin{eqnarray}\label{reflect4}
\tau_{-1}(-\tilde{x},t,-b+i\xi) = \sigma_{-1}(\tilde{x},t,b+i\xi).
\end{eqnarray}
\noindent Proceeding in this way, we find that all poles of the integrand in Eq.~(\ref{vint4}) must be embraced by the contour, the pole enumeration being defined by the following infinite chain of relations for their boundary values
\begin{eqnarray}\label{reflect5}
\tau_n(-\tilde{x},t,-b+i\xi) &=& \sigma_{n}(\tilde{x},t,b+i\xi), \nonumber\\
\tau_n(-\tilde{x},t,b+i\xi) &=& \sigma_{n-1}(\tilde{x},t,-b+i\xi),
\end{eqnarray}
\noindent for all integer $n.$ Note that for all $\tilde{x},$ each $\sigma$- or $\tau$-pole, except $\tau_0,$ belongs either to the upper or lower half of the $\tau$-plane. In fact, vanishing of the imaginary part of any of these poles for some $\tilde{x}$ would mean that there is a real trajectory $(X,Y)$ connecting the point $(\tilde{x},f(\tilde{x},\tau))$ with the observation point $(x,y)$, which is impossible since $(x,y)\in \Sigma,$ while all the image sources corresponding to the $\tau$- or $\sigma$ poles are outside of the channel. It is not difficult to see that ${\rm Im}\,\tau_0$ is negative (positive) when the observation point is at the right (left) wall. It follows then from Eqs.~(\ref{reflect3})--(\ref{reflect5}) that
\begin{eqnarray}\label{signs}&&
{\rm Im}\,\tau_n = {\rm Im}\,\sigma_n = \chi(n), \quad n\ne 0; \quad {\rm Im}\,\sigma_0 >0.
\end{eqnarray}
\noindent Thus, the integral kernel takes the form
\begin{eqnarray}\label{vint5}&&
\hspace{-0,5cm}K(\tilde{x},t,\zeta) \stackrel{\circ}{=} \frac{i}{2}\chi\left({\rm Im\,\tau_0} \right)\EuScript{M}(\tilde{x},t,\tau_0) - \frac{i}{2}\EuScript{M}(\tilde{x},t,\sigma_0) + \frac{i}{2}\sum_{n=-\infty}^{+\infty}\chi(n)\left\{\EuScript{M}(\tilde{x},t,\tau_n) - \EuScript{M}(\tilde{x},t,\sigma_n)\right\} \nonumber\\&&
- \frac{1}{16b}\int_{C_- \cup C_+}d\tau
\EuScript{M}(\tilde{x},t,\tau)\left(\cot\left\{\frac{\pi}{4b}\left[\zeta - g_{\tau}\right]\right\}\dot{g}_{\tau} - \tan\left\{\frac{\pi}{4b}\left[\zeta + \overline{g}_{\tau}\right]\right\}\dot{\overline{g}}_{\tau}\right)+ {\rm c.c.}
\end{eqnarray}
\noindent Integration over $\tau \in (-\infty,0)$ in the last term of this expression gives rise to a contribution which is $\stackrel{\circ}{=}0.$ Indeed, these $\tau$'s correspond to the integration over trajectories {\it before} they crossed the flame front (it is meant that these are trajectories of the burnt gas elements, continued to $\tau<0,$ not fresh gas elements). In other words, $Z_{\tau}\notin \Sigma$ for $\tau<0,$ and the Lemma allows us to omit this contribution. Furthermore, for all $\tau$'s corresponding to the remote semicircles, excluding a finite domain near some of the remote $\tau$- or $\sigma$-poles, the arguments of the tangent and cotangent have large imaginary parts of the same sign, hence, $\cot(\cdot) \to \pm i,$  $\tan(\cdot) \to \mp i,$ and the integrand becomes $\pm i \EuScript{M}(\tilde{x},t,\tau)(\dot{g}_{\tau} + \dot{\overline{g}}_{\tau}),$ plus an exponentially small remainder. The leading term is canceled by its counterpart from ``${\rm c.c.}$'' Indeed, using the definition of the function $\overline{g}_{\tau},$ we find
$$i\EuScript{M}(\tilde{x},t,\tau)(\dot{g}_{\tau} + \dot{\overline{g}}_{\tau}) + {\rm c.c.} = i\EuScript{M}(\tilde{x},t,\tau)\left(\frac{dg_{\tau}}{d\tau} + \frac{d\overline{g}_{\tau}}{d\tau}\right) - i\EuScript{M}(\tilde{x},t,\overline{\tau})\left(\frac{d\overline{g}_{\overline{\tau}}}{d\overline{\tau}} + \frac{dg_{\overline{\tau}}}{d\overline{\tau}}\right).$$ Changing the integration variable $\overline{\tau}\to \tau$ in the second term, and taking into account that $C_+, C_-$ are complex conjugates of each other shows that this expression is zero.
Thus, the integral over the expanding semicircles vanishes exponentially. On the other hand, if the sum in the first line of Eq.~(\ref{vint5}) converges, the contribution of the excluded domain also vanishes. In this case, therefore, the integral in Eq.~(\ref{vint5}) can be completely omitted.

In the course of the contour deformation, one may also encounter singularities of the functions $\EuScript{M}(\tilde{x},t,\tau),$ $g_{\tau}$ themselves. Under quite general assumptions about their analytic properties, it can be proved that these singularities contribute terms $\stackrel{\circ}{=}0.$ Below we consider only the case when $M(\tilde{x},t-\tau)$ is a meromorphic function of $\tau,$ {\it i.e.,} it is allowed to have any number of poles of arbitrary order. Any $k$th-order pole of the function $M(\tilde{x},t-\tau)$ becomes a $(k-1)$th-order pole of $\EuScript{M}(\tilde{x},\tau,t).$ Consider the case when $M(\tilde{x},t-\tau)$ has a second-order pole at some complex $\tau=\tau_*$ ($\tau_*$ is generally a function of $x$ and $t,$ but for brevity, we omit these arguments; also for definiteness, we assume that ${\rm Im}\,\tau_* >0$). Then $\EuScript{M}(\tilde{x},\tau,t)$ contains a term $m(\tilde{x},t)/(\tau - \tau_*),$ with some $m(\tilde{x},t)$ satisfying $m(-\tilde{x},t) = - m(\tilde{x},t).$ Reality of the function $\EuScript{M}(\tilde{x},\tau,t)$ implies that it has also a pole at $\tau = \overline{\tau}_*,$ and that the two poles have the same real residue. Crossing the pole $\tau_*$ by the contour $C_+$ adds to the right hand side of Eq.~(\ref{vint5}) a term
\begin{eqnarray}\label{addpole}&&
- \frac{2\pi i}{16b}m(\tilde{x},t)\left(\cot\left\{\frac{\pi}{4b}\left[\zeta - g_{\tau_*}\right]\right\}\dot{g}_{\tau_*} - \tan\left\{\frac{\pi}{4b}\left[\zeta + \overline{g}_{\tau_*}\right]\right\}\dot{\overline{g}}_{\tau_*}\right) + {\rm c.c.}
\end{eqnarray}
\noindent We see that this brings in an infinite sequence of singularities to the kernel $K(\tilde{x},t,\zeta)$ considered as a function of $\zeta.$ Namely, when $\zeta$ is close to $g_{\tau_*} + 4bl,$ $l \in Z,$ the new term reduces to a simple pole
$$- \frac{im(\tilde{x},t)}{2}\frac{\dot{g}_{\tau_*}}{\zeta - g_{\tau_*} - 4bl}\,.$$ On the other hand, since the poles of the cotangent are in one-to-one correspondence with integer numbers, there is $n$ such that $g_{\tau_n} = \zeta - 4bl.$ It follows from $\zeta \to g_{\tau_*} + 4bl$ that $\tau_n \to \tau_*.$ In particular, ${\rm Im}\,\tau_n >0,$ therefore, the $n$th term of the series on the right of Eq.~(\ref{vint5}) contributes $+(i/2)m(\tilde{x},t)/(\tau_n - \tau_*).$ Furthermore,  $g_{\tau_n} = g_{\tau_*} + \dot{g}_{\tau_*}(\tau_n - \tau_*) + o(\tau_n - \tau_*),$ or, $g_{\tau_*} = \zeta - 4bl - \dot{g}_{\tau_*}(\tau_n - \tau_*) + o(\tau_n - \tau_*),$ so the sum of the two contributions is
$$- \frac{im(\tilde{x},t)}{2}\frac{\dot{g}_{\tau_*}}{\zeta - g_{\tau_*} - 4bl} + \frac{i}{2}\frac{m(\tilde{x},t)}{\tau_n - \tau_*} = O(1)\,.$$ A similar consideration shows that the same is true of the pole $\overline{\tau}_*,$ and of the $\sigma$-contributions (the latter cancel the poles of tangent in the expression (\ref{addpole})). When $l$ runs over all integers, so does $n,$ and we see that all pole contributions to the right hand side of Eq.~(\ref{vint5}) cancel. Differentiating with respect to $\tau_*,$ it is not difficult to see that poles with $k\geqslant 3$ do not contribute either.

Next, let $M(\tilde{x},t-\tau)$ have simple poles at $\tau_*,$ $\overline{\tau}_*.$ These poles contribute to the function $\EuScript{M}(\tilde{x},\tau,t)$ a term $m(\tilde{x},t)\ln\left\{(\tau - \tau_*)(\tau - \overline{\tau}_*)\right\},$ with some $m(\tilde{x},t)$ having the same properties as before. Crossing the logarithmic singularities leads to the $2\pi$ jump in $\arg(\ln(\cdot))$ for all points of the contour $C_- \cup C_+,$ located at one side of the point $\tau_*$ or $\overline{\tau}_*,$ so the pole contribution to the integral kernel reads
\begin{eqnarray}&&\label{addlog}
\frac{\pi i}{8b}m(\tilde{x},t)\int_{\overline{\tau}_*}^{\tau_*}d\tau
\left(\cot\left\{\frac{\pi}{4b}\left[\zeta - g_{\tau}\right]\right\}\dot{g}_{\tau} - \tan\left\{\frac{\pi}{4b}\left[\zeta + \overline{g}_{\tau}\right]\right\}\dot{\overline{g}}_{\tau}\right) + {\rm c.c.} \nonumber\\&& = - \frac{i}{2}m(\tilde{x},t)\left[\ln\sin\left\{\frac{\pi}{4b}\left[\zeta - g_{\tau}\right]\right\} - \ln\cos\left\{\frac{\pi}{4b}\left[\zeta + \overline{g}_{\tau}\right]\right\}\right]_{\overline{\tau}_*}^{\tau_*} + {\rm c.c.}
\end{eqnarray}
\noindent (we may assume that the contour avoids singularities of the integrand, because they are isolated. Then all singularities of the integral are associated with the contour endpoints). For $\zeta$ close to $g_{\tau_*} + 4bl,$ $l \in Z,$ the singularity is
$$- \frac{im(\tilde{x},t)}{2}\ln(\zeta - g_{\tau_*} - 4bl).$$ It is compensated by $+i/2\ln(\tau_n - \tau_*)$ coming from the $n$-th term of the series in Eq.~(\ref{vint5}), where $n$ is such that $g_{\tau_n} = \zeta - 4bl.$ Again, cancelation of singularities takes place for all $n.$

Thus, we arrive at the conclusion that the contour deformation ``wipes out'' all poles of $M,$ so that no pole or logarithmic singularity is left in the kernel $K.$

Now, this result can be used to prove that any pole or branch singularity in $g_{\tau}$ gives rise only to terms $\stackrel{\circ}{=}0.$ Namely, if $g_{\tau}$ has a pole at $\tau_*,$ then its contribution to the right hand side of Eq.~(\ref{vint5}) reads
\begin{eqnarray}&&\label{addlogg}
- \frac{\EuScript{M}(\tilde{x},t,\tau_*)}{16b}\int_{C_*}d\tau
\left(\cot\left\{\frac{\pi}{4b}\left[\zeta - g_{\tau}\right]\right\}\dot{g}_{\tau} - \tan\left\{\frac{\pi}{4b}\left[\zeta + \overline{g}_{\tau}\right]\right\}\dot{\overline{g}}_{\tau}\right)+ {\rm c.c.},
\end{eqnarray}
\noindent
where $C_*$ is a contour surrounding $\tau_*,$ which can be taken as small as desired. Furthermore, if there is a finite cut in the $\tau$-plane, connecting  branch points of $g_{\tau},$ it brings in a term
\begin{eqnarray}&&\label{addroot}
- \frac{1}{16b}\int_{C_0}d\tau
\EuScript{M}(\tilde{x},t,\tau)\left(\cot\left\{\frac{\pi}{4b}\left[\zeta - g_{\tau}\right]\right\}\dot{g}_{\tau} - \tan\left\{\frac{\pi}{4b}\left[\zeta + \overline{g}_{\tau}\right]\right\}\dot{\overline{g}}_{\tau}\right) + {\rm c.c.},
\end{eqnarray}
\noindent where $C_0$ is a contour embracing the cut. Both expressions (\ref{addlogg}) and (\ref{addroot}) contain logarithmic singularities with respect to $\zeta.$ But the right hand side of Eq.~(\ref{vint5}) has just been proved to be free of such contributions, so the new singularities have no counterterms. At the same time, the burnt gas velocity is bounded, as are all the irrotational contributions omitted in the course of its simplification [Cf. condition b)]. Hence, the singular point and the cut must be such that $g_{\tau}\notin g(\Sigma)$ for $\tau=\tau_*$ or $\tau\in C_0.$ Noting also that since $\EuScript{M}$ is odd in $\tilde{x},$ so is $[\EuScript{M}],$ and applying the Lemma, we see that the above expressions $\stackrel{\circ}{=}0$ indeed.

To summarize, under the assumption of convergence of the series over $\tau$- and $\sigma$-poles, the function (\ref{kernel}) can be written
\begin{eqnarray}\label{vint6}&&
\hspace{-0,4cm}K(\tilde{x},t,\zeta) \stackrel{\circ}{=} K_{\rm c}(\tilde{x},t,\zeta) + \overline{K_{\rm c}(\tilde{x},t,\zeta)}, \\&&
\hspace{-0,4cm} K_{\rm c}(\tilde{x},t,\zeta) = \frac{i}{2}\chi\left({\rm Im\,\tau_0} \right)\EuScript{M}(\tilde{x},t,\tau_0) - \frac{i}{2}\EuScript{M}(\tilde{x},t,\sigma_0) + \frac{i}{2}\sum_{n=-\infty}^{+\infty}\chi(n)\left\{\EuScript{M}(\tilde{x},t,\tau_n) - \EuScript{M}(\tilde{x},t,\sigma_n)\right\},\nonumber
\end{eqnarray}
\noindent it being understood that all pole or logarithmic singularities, if any,  have been removed from $\EuScript{M}$ (for brevity, we do not introduce special designation for this truncated function).

To conclude this section, let us show that the velocity field
$$\omega^v(z) = \frac{\slashed{\partial}}{2}\int_{-B}^{+B}d\tilde{x}K(\tilde{x},t,g(z))$$
satisfies
$\Omega^v(-b+i\xi) = \Omega^v(b+i\xi)$ for all $\xi$ downstream, where $\Omega^v(\zeta) = \omega^v(g^{-1}(\zeta))\frac{\displaystyle dg^{-1}(\zeta)}{\displaystyle d\zeta}.$ Note that since $\EuScript{M}(\tilde{x},t,\tau)$ is real for $\tau$ real, differentiation of the $\chi$-function in the first term in $K_{\rm c}$ yields a purely imaginary quantity canceled by its counterpart from $\overline{K}_{\rm c}.$ This means that when calculating the velocity field, $K_{\rm c}$ depends in effect on $z = x + iy,$ while its complex conjugate, on $\overline{z}= x - iy,$ enabling us to apply the rule (\ref{diff}) which gives
$$\Omega^v(\zeta) = \frac{\partial}{\partial\zeta}\int_{-B}^{+B}d\tilde{x}K_{\rm c}(\tilde{x},t,\zeta).$$ Therefore, it is sufficient to prove that
\begin{eqnarray}\label{toprove}
\int_{-B}^{+B}d\tilde{x}K_{\rm c}(t,\tilde{x},-b+i\xi) = \int_{-B}^{+B}d\tilde{x}K_{\rm c}(t,\tilde{x},b+i\xi).
\end{eqnarray}
\noindent
Using Eq.~(\ref{reflect5}) we write
\begin{eqnarray}
\sum_{n=-\infty}^{+\infty}&&\hspace{-0,3cm}\chi(n)\left\{
\EuScript{M}(\tilde{x},t,\tau_{n}(\tilde{x},t,-b+i\xi)) - \EuScript{M}(\tilde{x},t,\sigma_{n}(\tilde{x},t,-b+i\xi))\right\} \nonumber\\ = \sum_{n=-\infty}^{+\infty}&&\hspace{-0,3cm}\chi(n)\left\{
\EuScript{M}(\tilde{x},t,\sigma_{n}(-\tilde{x},t,b+i\xi)) - \EuScript{M}(\tilde{x},t,\tau_{n}(-\tilde{x},t,b+i\xi))\right\} \nonumber\\&& + \EuScript{M}(\tilde{x},t,\tau_{0}(-\tilde{x},t,b+i\xi)) +\EuScript{M}(\tilde{x},t,\tau_{1}(-\tilde{x},t,b+i\xi)).\nonumber
\end{eqnarray}
\noindent Similarly,
\begin{eqnarray}&&
\EuScript{M}(\tilde{x},t,\tau_0(\tilde{x},t,-b+i\xi)) - \EuScript{M}(\tilde{x},t,\sigma_0(\tilde{x},t,-b+i\xi)) \nonumber\\&&=
\EuScript{M}(\tilde{x},t,\sigma_0(-\tilde{x},t,b+i\xi)) - \EuScript{M}(\tilde{x},t,\tau_1(-\tilde{x},t,b+i\xi)).\nonumber
\end{eqnarray}
\noindent Adding these identities, changing the integration variable $\tilde{x}\to - \tilde{x}$ on the left of Eq.~(\ref{toprove}), and taking into account that $\EuScript{M}(-\tilde{x},t,\zeta) = - \EuScript{M}(\tilde{x},t,\zeta)$ proves this equality. In fact, it can be shown that $\Omega^v(\pm b + i\xi)=0,$ but we will not need this. Thus, the property $\Omega^v(-b+i\xi) = \Omega^v(b+i\xi)$ is a direct consequence of the chain relations (\ref{reflect5}) and antisymmetry of $\EuScript{M}$ with respect to $\tilde{x}.$ This fact will be used in the next section.

\subsection{Evaluation of rotational velocity component near the flame front}\label{nearfront}

Let us now specialize to the case when the observation point $z$ is close to the flame front. In this case, the $\tau_0$-pole crosses the real axis of the complex $\tau$-plane near its origin. For sufficiently small $|\tau|,$ one has,
according to the definition of particle trajectory,
\begin{eqnarray}\label{traject}
X(\tilde{x},t,t-\tau) & = & \tilde{x} + w_+(\tilde{x},t)\tau,\nonumber\\ Y(\tilde{x},t,t-\tau) & = &
f(\tilde{x},t-\tau) + u_+(\tilde{x},t)\tau.
\end{eqnarray} Expanding also $f(\tilde{x},t-\tau)$ to the first order
in $\tau,$ we find for the complex trajectory
$$Z_{\tau} = \tilde{z} + [w_+(\tilde{x},t) + i\hat{u}_+(\tilde{x},t)]\tau,$$ so that the equation $g(Z_{\tau}) = \zeta$ becomes
\begin{eqnarray}\label{point2}&&
g(\tilde{z}) + \frac{dg}{dz}(\tilde{z})[w_+(\tilde{x},t) + i\hat{u}_+(\tilde{x},t)]\tau = \zeta.
\end{eqnarray} The value of the rotational velocity component is given by integrating $K_0(\zeta,\tilde{x},t)$ over $\tilde{x}.$ If $\tilde{x}$ is sufficiently close to the point determined by Eq.~(\ref{point1}), then Eq.~(\ref{point2}) gives
\begin{eqnarray}\label{tau0}
\tau_0(\tilde{x},t,\zeta) = \frac{\zeta - g(\tilde{z})}{\tilde{\gamma}[w_+(\tilde{x},t) + i\hat{u}_+(\tilde{x},t)]}\,, \quad \tilde{\gamma} \equiv \frac{dg}{dz}(\tilde{z})\,.
\end{eqnarray}\noindent
For other $\tilde{x}$'s which are not close to the point satisfying (\ref{point1}), $\tau_0$ has different form depending on the specific structure of higher-order terms with respect to $\tau$ in Eq.~(\ref{traject}). However, for such $\tilde{x}$'s the pole $\tau_0$ is far from the real axis. On the other hand, its exact position in the complex plane is inessential for the purpose of calculating the rotational velocity component, because any displacement of this pole adds to velocity terms $\stackrel{\circ}{=}0,$ as long as the pole does not cross the real axis during this displacement. More precisely, it was shown in Sec.~\ref{generalst} that any pole or logarithmic singularities are removed from $\EuScript{M}$ by the contour deformation, hence, displacements of $\tau_0$ in Eq.~(\ref{vint6}) respect conditions a) and b). However, changing location of $\tau_0$ generally violates c). To ensure that this condition is still met, it is sufficient to fulfil the chain relations (\ref{reflect5}). Indeed, we saw at the end of the preceding section that the latter entail equality $\Omega^v(-b+i\xi) = \Omega^v(b+i\xi)$ for the rotational velocity component. Therefore, the difference of the functions $\Omega^v(\zeta)$ corresponding to this component before and after the displacement will satisfy c). Thus, we shift the pole $\tau_0$ so that its new position is given by Eq.~(\ref{tau0}) for all $\tilde{x},$ and accordingly, shift the infinite sequence of image poles to the new positions given by
\begin{eqnarray}\label{tausigma}
\tau_n(\tilde{x},t,\zeta) = \frac{\zeta - 4 b n - g(\tilde{z})}{\tilde{\gamma}[w_+(\tilde{x},t) + i\hat{u}_+(\tilde{x},t)]}\,, \quad \sigma_n(\tilde{x},t,\zeta) = \frac{4 b n + 2b - \zeta - \overline{g(\tilde{z})}}{\overline{\tilde{\gamma}}[w_+(\tilde{x},t) - i\hat{u}_+(\tilde{x},t)]}\,.
\end{eqnarray}\noindent Let us check that Eq.~(\ref{reflect5}) holds true for the new pole location. Taking into account that $-\tilde{x} + i f(-\tilde{x},t) = -\overline{\tilde{z}},$ one finds
\begin{eqnarray}
\tau_n(-\tilde{x},t,-b + i\xi) &=& \frac{-b + i\xi - 4 b n - g(-\overline{\tilde{z}})}{\displaystyle\frac{dg}{dz}(-\overline{\tilde{z}})[w_+(-\tilde{x},t) + i\hat{u}_+(-\tilde{x},t)]}\,, \nonumber\\
\sigma_n(\tilde{x},t,b + i\xi) &=& \frac{4 b n + b - i\xi - \overline{g(\tilde{z})}}{\displaystyle\overline{\frac{dg}{dz}(\tilde{z})}[w_+(\tilde{x},t) - i\hat{u}_+(\tilde{x},t)]}\,.\nonumber
\end{eqnarray}\noindent
Differentiating Eq.~(\ref{reflect}) with respect to $x$ gives
\begin{eqnarray}\label{reflectd}
\frac{dg}{dz}(-\bar{z}) = \overline{\frac{dg}{dz}(z)}\,,
\end{eqnarray}
\noindent which together with Eqs.~(\ref{reflect}), (\ref{reflect1}) prove the first of the two infinite sequences of relations (\ref{reflect5}). The other is proved similarly. Thus, under the assumption of convergence of the series, rotational component of the burnt gas velocity at the flame front takes the form
\begin{eqnarray}\label{vintf}
\omega^v = &&\hspace{-0,3cm}-\frac{\slashed{\partial}}{2}\,{\rm Im}\int_{-B}^{+B}d\tilde{x}[ \chi\left({\rm Im\,\tau_0} \right)\EuScript{M}(\tilde{x},t,\tau_0) - \EuScript{M}(\tilde{x},t,\sigma_0) \nonumber\\&& + \sum_{n=-\infty}^{+\infty}\chi(n)\left\{\EuScript{M}(\tilde{x},t,\tau_n) - \EuScript{M}(\tilde{x},t,\sigma_n)\right\}],
\end{eqnarray}
\noindent it being understood that $y$ is set equal to $f(x,t)$ after the differentiation.

\subsection{The master equation}\label{mastereq}

We can finally write the main integro-differential equation relating the on-shell value of the fresh-gas velocity and the flame front position. As we proved in the preceding section, the irrotational component of the burnt gas velocity, $\bm{v}^p,$ which was gradually removed from the burnt-gas velocity downstream, satisfies conditions a)--c), so that the corresponding function $\Omega^p(\zeta) = \omega^p(g^{-1}(\zeta))\frac{\displaystyle dg^{-1}(\zeta)}{\displaystyle d\zeta}$ obeys Eq.~(\ref{dispers1}). This relation can be put in a more compact form by introducing operator $\hat{\EuScript{H}\,}$ defined by
\begin{eqnarray}\label{hilbert}
\left(\hat{\EuScript{H}\,}a\right)(x) = \frac{g'(z_+)}{2b}\fint_{-B}^{+B}d\tilde{x} \cot\left\{\frac{\pi}{2b}[g(\tilde{z}) - g(z_+)]\right\}a(\tilde{x}),
\end{eqnarray}
\noindent where $a(x)$ is assumed to be continuous function having zero average across the channel, {\it i.e.,}
\begin{eqnarray}\label{average}
\langle a \rangle \equiv \frac{1}{2B}\int_{-B}^{+B}d\tilde{x} a(\tilde{x}) = 0\,.
\end{eqnarray}
\noindent It is proved in Appendix that under this condition, $\hat{\EuScript{H}\,}$ satisfies the usual identity
\begin{eqnarray}\label{identity}
\hat{\EuScript{H}\,}^2 = - 1.
\end{eqnarray}
\noindent
The dispersion relation for $\omega^p$ thus becomes
\begin{eqnarray}\label{dispers2}
\left(1 + i\hat{\EuScript{H}\,}\right)\left(\Omega^p_+\right)' = 0.
\end{eqnarray}
\noindent A calculation quite similar to that carried out in Sec.~\ref{dispersrel} shows that the fresh-gas velocity, being irrotational by virtue of the Thomson theorem, satisfies
\begin{eqnarray}\label{dispers3}
\left(1 - i\hat{\EuScript{H}\,}\right)\left(\Omega_-\right)' = 0.
\end{eqnarray}
\noindent The proof that the function $\Omega(\zeta) = \omega(g^{-1}(\zeta))\frac{\displaystyle dg^{-1}(\zeta)}{\displaystyle d\zeta}$ corresponding to the fresh-gas velocity fulfils a)--c) (with the requirement of boundedness applied upstream) is the same as for $\Omega_0$ given in Sec.~\ref{generalst}. Let $[\bm{v}]$ denote the jump of the gas velocity across
the flame front, $[\bm{v}] = \bm{v}(x,f(x,t)+0) -
\bm{v}(x,f(x,t)-0).$ Then the master equation for $\omega_-,f$
is obtained by substituting
$$\Omega^p_+ =  \frac{[\omega] - \omega^v_+}{\gamma_+} + \Omega_-\,, \quad \gamma_+ \equiv \gamma(z_+) = \frac{dg}{dz}(z_+)\,,  $$ in Eq.~(\ref{dispers2}), and using Eqs.~(\ref{vintf}), (\ref{dispers3})
\begin{eqnarray}\label{master}
2\left(\frac{\omega_-}{\gamma_+}\right)' + \left(1 + i\hat{\EuScript{H}\,}\right)\left\{ \frac{[\omega]}{\gamma_+}  + \frac{\slashed{\partial}}{2\gamma_+}{\rm Im}\int_{-B}^{+B}d\tilde{x}\left[\phantom{\sum_{n=-\infty}^{+\infty}}\hspace{-0,9cm}\chi\left({\rm Im\,\tau_0} \right)\EuScript{M}(\tilde{x},t,\tau_0) - \EuScript{M}(\tilde{x},t,\sigma_0) \right.\right.\nonumber\\\left.\left. + \sum_{n=-\infty}^{+\infty}\chi(n)\left\{\EuScript{M}(\tilde{x},t,\tau_n) - \EuScript{M}(\tilde{x},t,\sigma_n)\right\}\right]\right\}' = 0.
\end{eqnarray}
\noindent By virtue of the identity (\ref{identity}), solutions to this equation automatically satisfy Eq.~(\ref{dispers3}).
The quantities $\sigma_+,$ $\hat{v}^n_+$ determining the memory kernel, as well as the velocity jumps appearing in this equation, are all known functionals of the on-shell fresh gas velocity \citep{matalon,pelce}. For instance, in the simplest case of zero-thickness flame fronts one has
\begin{eqnarray}\label{jumps}
\hat{v}^n_+ &=& \theta\,, \quad [u] = \frac{\theta - 1}{N}\ ,
\quad [w] = - f'\frac{\theta - 1}{N}\,, \\
\sigma_+ &=& - \frac{\theta - 1}{\theta N}\left\{\frac{\EuScript{D}w_-}{\EuScript{D}t} +
f' \frac{\EuScript{D}u_-}{\EuScript{D}t} + \frac{1}{N}\frac{\EuScript{D}f'}{\EuScript{D}t}
\right\}\,,\label{vorticity}
\end{eqnarray}
\noindent where
$$\frac{\EuScript{D}}{\EuScript{D}t} \equiv \frac{\partial}{\partial t} + \left(w_- +
\frac{f'}{N}\right)\frac{\partial}{\partial x}\,.$$ In addition to that, the local consumption rate in the flame is given by the so-called evolution equation
\begin{eqnarray}\label{evolutiongen}
(\hat{\bm{v}}_-, \bm{n}) = 1 + S(u_-,w_-,f')\,,
\end{eqnarray}
\noindent where $S$ is a known functional of its arguments,
proportional to the cutoff wavelength $\lambda_{\rm c}.$ Together with this equation, the complex Eq.~(\ref{master}) describes unsteady flame propagation in curved channels  in the most general form.

Let us consider the special case of straight channel. Setting $g(z)=z,$ $\gamma_+=1,$ expressions (\ref{tausigma}) can be rewritten as
\begin{eqnarray}
\tau_n(\tilde{x},t,z) = - i\frac{\hat{\omega}_+}{\hat{v}^2_+}(z - [\tilde{z} + 4 b n])\,, \quad \sigma_n(-\tilde{x},t,z) = - i\frac{\hat{\omega}_+}{\hat{v}^2_+}(z - [\tilde{z} + 4 b n + 2b])\,,\nonumber
\end{eqnarray}\noindent where $\hat{\omega}_+ = \hat{u}_+ + iw_+,$ $\hat{v}^2_+ = w^2_+ + \hat{u}^2_+.$ It is seen that the two infinite sequences of poles span unique sequence $$T_n(\tilde{x},t,z) = - i\frac{\hat{\omega}_+}{\hat{v}^2_+}(z - [\tilde{z} + 2 b n])\,,\quad n \in Z.$$ Furthermore, if we extend the $\tilde{x}$-domain from $[-b,b]$ to $(-\infty,\infty),$ and change $\tilde{x}\to - \tilde{x}$ in the sum over $\sigma$-poles, the $\tilde{x}$-integral in Eq.~(\ref{master}) becomes, taking into account oddness of $\EuScript{M}$ and Eq.~(\ref{signs}),
$$\int_{-\infty}^{+\infty}d\tilde{x}\chi\left({\rm Im}\,T \right)\EuScript{M}(\tilde{x},t,T),$$ where
$$T = T(\tilde{x},t,z) = - i\frac{\hat{\omega}_+}{\hat{v}^2_+}(z - \tilde{z}).$$ Introducing vector $\bm{r} = (x-\tilde{x},f(x,t)-f(\tilde{x},t)),$ this can also be written as $$T = \frac{(\hat{\bm{v}}_+,\bm{r})}{\hat{v}^2_+} + i\chi\left({\rm Im}\, T\right)\frac{r}{\hat{v}_+}\sqrt{1 - \frac{(\hat{\bm{v}}_+,\bm{r})^2}{\hat{v}^2_+r^2}}.$$ Recalling the definition (\ref{eumdef}), the master equation thus reads
\begin{eqnarray}
2\left(\omega_-\right)' + \left(1 + i\hat{\EuScript{H}\,}\right)\left\{ [\omega] + \frac{i\slashed{\partial}}{4}\int_{-\infty}^{+\infty}d\tilde{x}\int_{\tau_-}^{\tau_+} d\tau M(\tilde{x},t-\tau)\right\}' = 0,\nonumber
\end{eqnarray}
\noindent where
$$\tau_{\pm} = \frac{(\hat{\bm{v}}_+,\bm{r})}{\hat{v}^2_+} \pm i\frac{r}{\hat{v}_+}\sqrt{1 - \frac{(\hat{\bm{v}}_+,\bm{r})^2}{\hat{v}^2_+r^2}}\,.$$
This equation coincides with Eq.~(5.5) of Ref.~\cite{jerk2} where it was shown, in particular, that complemented by the evolution equation, it contains all theoretical results on flame dynamics established so far, including the linear equation describing Darrieus-Landau instability of planar flames, and the
nonlinear Sivashinsky-Clavin equation \cite{sivclav} for flames with weak gas expansion.

To ensure convergence of the improper integral over $\tilde{x},$ it was regularized in Ref.~\cite{jerk2} by introducing the factor $e^{-\mu r}$ with sufficiently large positive $\mu,$ the rotational velocity component being understood as the analytic continuation of the regularized expression to $\mu = 0.$ A similar but more physical regularization needed generally to ensure convergence of the series in Eq.~(\ref{master}) will be introduced in the next section.

\section{Steady flame in a curved channel}\label{steady}

\begin{figure}
\centering
\includegraphics[width=0.4\textwidth]{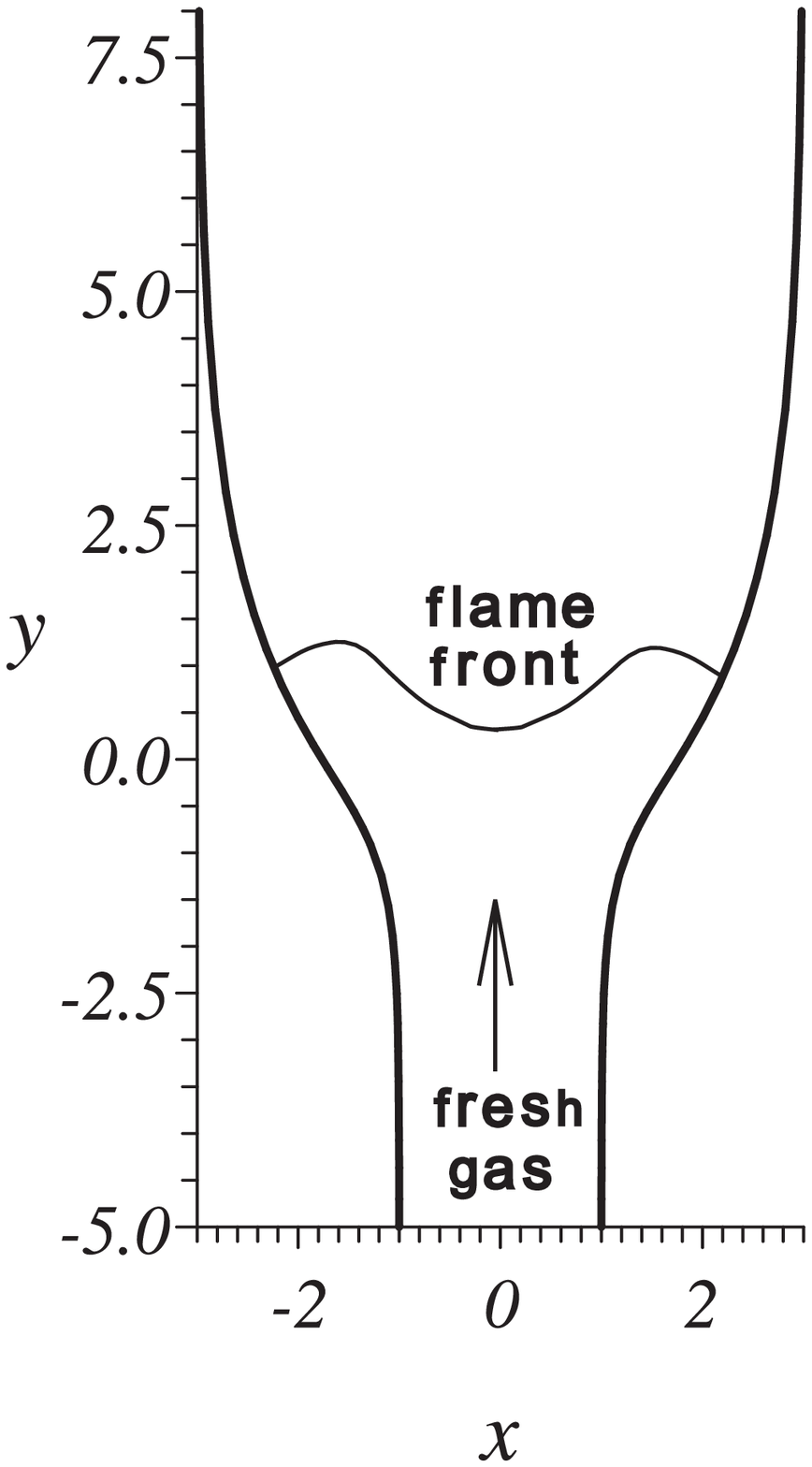}
\caption{}\label{fig3}
\end{figure}

A case of special interest for application of the developed theory is the steady flame propagation. To consider it in a curved channel, we have to assume that there is an incoming flow of fresh gas, large enough to hold the front at a given position in the channel. Take, for instance, a channel shaped bottlewise as shown on Fig.~\ref{fig3}. The map from the auxiliary to the physical plane is given by
\begin{eqnarray}\label{bottle}
g^{-1}(\zeta) = \zeta + \frac{2i}{\pi}(B_{\infty} - b)\ln\left\{1 + \exp\left(-\frac{i\pi \zeta}{2b}\right)\right\}\,,
\end{eqnarray}
\noindent
where $b$ and $B_{\infty}$ are the channel half-width in the far up- and downstream, respectively. If the incoming  flow velocity exceeds the relative curved flame speed in the straight channel of width $b,$ and $B_{\infty}$ is large enough (such that the flow velocity at $z=+\infty$ is less than the normal flame speed), then the flame will be stabilized somewhere in the convergent part of the bottleneck. It is easy to see that the master equation (\ref{master}) still applies to the case under consideration despite different initial conditions. In fact, it is evident from the derivation of the expression (\ref{vintf}) for the rotational velocity component that it has the same form whatever the net gas flow. On the other hand, the dispersion relations for the irrotational velocity components involve differentiated functions $\Omega = \omega/(dg/dz)$ which are nothing but the flow velocities in the auxiliary plane. Therefore, a constant shift of the velocities does not change these relations either. In the steady case, the memory kernel is time-independent, $M(\tilde{x},t)\equiv M(\tilde{x}),$ so that Eq.~(\ref{master}) would take the form
\begin{eqnarray}\label{masters}
2\left(\frac{\omega_-}{\gamma_+}\right)' + \left(1 + i\hat{\EuScript{H}\,}\right)&&\hspace{-0,3cm}\left\{\frac{[\omega]}{\gamma_+}  - \frac{\slashed{\partial}}{2\gamma_+}\int_{-B}^{+B}d\tilde{x}M(\tilde{x})\right.\nonumber\\&&\left. \times\left[|{\rm Im}\,\tau_0| - {\rm Im}\,\sigma_0 + {\rm Im}\sum_{n=-\infty}^{+\infty}\chi(n)(\tau_n - \sigma_n)\right]\right\}' = 0.
\end{eqnarray}
\noindent However, as written this equation has only a formal meaning, because the infinite series in the braces does not converge, as can be seen using the explicit formulas (\ref{tausigma}). One might think that this complication is not essential, since each term in this sum is a linear function of $\zeta = g(z).$ Indeed, if the term-by-term differentiation of the series were allowed, then $\slashed{\partial}g(z) = 2dg/dz$ taken on-shell would just cancel $\gamma_+$ in the denominator, leaving us with an $x$-independent expression which falls off from the equation upon the subsequent $x$-differentiation. But this operation is not allowed, because Eq.~(\ref{master}) was obtained under the assumption of convergence of the series. A physical way to resolve this ambiguity is to recall that the gas viscosity, though usually negligible in practice, never vanishes in principle. This fact suggests the following natural regularization of the series. In the case of a steady viscous flow, Eq.~(\ref{flow2}) is replaced by $$(\bm{v},\bm{\nabla})\sigma = \nu\triangle\sigma,$$ where $\nu$ is the kinematic viscosity. It follows from this equation that for sufficiently small $\nu,$ vorticity decays along the stream lines roughly as $\sigma \sim \exp(-\nu\tau/b^2).$ Since $M(\tilde{x})$ is proportional to $\sigma,$ this means that the definition (\ref{eumdef}) of the function $\EuScript{M}(\tilde{x},t,\tau)$ is to be replaced by
$$\EuScript{M}(\tilde{x},t,\tau) = \lim\limits_{\varepsilon\to 0}\int\limits_{\tau}^{+\infty}d\tau_1 M(\tilde{x},t-\tau_1)e^{-\varepsilon\tau_1}$$ (according to the discussion at the end of Sec.~\ref{integrep}, this integral converges for $\tau_1\to \infty,$ so the limit does exist). The parameter $\varepsilon>0$ here is generally a function of $\tilde{x},$ but for brevity, we suppress this dependence. We need the imaginary part of $\EuScript{M},$ which can be written as
\begin{eqnarray}\label{mreg}
{\rm Im}\,\EuScript{M}(\tilde{x},t,\tau) = \frac{1}{2i}\lim\limits_{\varepsilon\to 0}\int\limits_{\tau}^{\overline{\tau}}d\tau_1 M(\tilde{x},t-\tau_1)e^{-\varepsilon\tau_1}\,.
\end{eqnarray}
\noindent
In the steady case, this becomes
$${\rm Im}\,\EuScript{M}(\tilde{x},\tau) = \frac{M(\tilde{x})}{2i}\lim\limits_{\varepsilon\to 0}\frac{e^{-\varepsilon\tau} - e^{-\varepsilon\overline{\tau}}}{\varepsilon}\,,$$
implying that the series in Eq.~(\ref{masters}) is to be understood as
$$\frac{1}{2i}\lim\limits_{\varepsilon\to 0}\frac{1}{\varepsilon}\sum_{n=-\infty}^{+\infty}\chi(n)
\left\{- e^{-\varepsilon\tau_n} + e^{-\varepsilon\overline{\tau_n}} + e^{-\varepsilon\sigma_n} - e^{-\varepsilon\overline{\sigma_n}}\right\}.$$ Substituting Eq.~(\ref{tausigma}), summing the geometric series, and taking the limit $\varepsilon\to 0$ gives
$${\rm Im}\left\{-\sigma_0 + \sum_{n=-\infty}^{+\infty}\chi(n)(\tau_n - \sigma_n)\right\} = \frac{1}{8bi}\left\{\frac{[\zeta - g(\tilde{z})]^2}{D} + \frac{[\zeta + \overline{g(\tilde{z})}]^2}{\overline{D}} - {\rm c.c.}\right\},$$ where $D = D(\tilde{x})\equiv \tilde{\gamma}[w_+(\tilde{x}) + iu_+(\tilde{x})].$ It should be mentioned that the limits of the sums over $\tau$- and $\sigma$-poles do not exist separately. For instance, one has
\begin{eqnarray}\label{series}
\sum_{n=-\infty}^{+\infty}\chi(n)e^{-\varepsilon\sigma_n} = \exp\left\{\displaystyle\frac{\varepsilon}{\overline{D}} [\zeta + \overline{g(\tilde{z})}]\right\}\frac{2\cosh\left\{\displaystyle\frac{2\varepsilon b}{\overline{D}}\right\}}{\exp\left\{\displaystyle\frac{4b\varepsilon}{\overline{D}}\right\} - 1}\,.
\end{eqnarray}
\noindent
It is evident that the imaginary part of this expression is $O(1/\varepsilon).$ It is canceled by a similar contribution coming from the sum over $\tau$-poles, as is the next $O(1)$-term. In effect, the exponents in the numerator are to be expanded to the second order in $\varepsilon,$ which is the reason why the resulting expression turns out to be quadratic in $\zeta.$ Putting it into Eq.~(\ref{masters}), and performing $\slashed{\partial}$-differentiation with the help of the rule (\ref{diff}) yields
\begin{eqnarray}
2\left(\frac{\omega_-}{\gamma_+}\right)' + \left(1 + i\hat{\EuScript{H}\,}\right)\left\{\frac{[\omega]}{\gamma_+}  - \int_{-B}^{+B}d\tilde{x}M(\tilde{x})\left[\frac{\chi_+\left({\rm Im}\,\tau_0\right)}{2iD} + \frac{g(z_+)}{4ib}\left(\frac{1}{D} + \frac{1}{\overline{D}}\right)\right]\right\}' = 0.\nonumber
\end{eqnarray}
\noindent Finally, noting that $\chi_+\left({\rm Im}\,\tau_0\right) = \chi(\tilde{x} - x),$ the $x$-derivative of the first term in the integrand removes the $\tilde{x}$-integral, while changing $\tilde{x}\to -\tilde{x}$ in the second and taking into account that $\overline{D(-\tilde{x})} = -D(\tilde{x})$ brings the master equation to the form
\begin{eqnarray}\label{masters1}
2\left(\frac{\omega_-}{\gamma_+}\right)' + \left(1 + i\hat{\EuScript{H}\,}\right)\left\{\left(\frac{[\omega]}{\gamma_+}\right)'  - \frac{M(x)\omega_+(x)}{\gamma_+ v^2_+(x)} + \frac{g'(z_+)}{2b}\int_{-B}^{+B}d\tilde{x}\frac{M(\tilde{x})\omega_+(\tilde{x})}{\tilde{\gamma} v^2_+(\tilde{x})}\right\} = 0.
\end{eqnarray}
\noindent The last term in this equation, resulting from the infinite series over $\tau$- and $\sigma$-poles, ensures that the expression in the braces has zero average across the channel, as it should be according to the definition of the operator $\hat{\EuScript{H}\,}.$ Indeed, it follows from $g(z_+)|_{x=\pm B} = \pm b + i\xi$ that the average of the last two terms in the braces vanishes, while
$$\left\langle \left(\frac{[\omega]}{\gamma_+}\right)'\right\rangle = \frac{1}{2B}\left.\frac{[\omega]}{\gamma_+}\right|_{-B}^{+B}.$$ Evidently, the quantity $1/\gamma_+$ taken at the channel wall represents vector normal to the wall at the front end-points. Since at these points gas velocity is tangent to the wall, so is its jump, which means that $[\omega]/\gamma_+$ is real there. On the other hand, the values of $[\omega]/\gamma_+$ at the left and the right wall are complex conjugates of each other, as is seen from Eqs.~(\ref{reflect1}), (\ref{reflectd}), thus proving that the average of the first term in the braces is also zero.

Let us consider flame propagation in a straight channel in more detail. The master equation reduces in this case to
\begin{eqnarray}\label{masterst}
2\left(\omega_-\right)' + \left(1 + i\hat{\EuScript{H}\,}\right)\left\{[\omega]'  - \frac{M(x)\omega_+(x)}{v^2_+(x)} + \frac{z'_+}{2b}\int_{-b}^{+b}d\tilde{x}\frac{M(\tilde{x})\omega_+(\tilde{x})}{v^2_+(\tilde{x})}\right\} = 0,
\end{eqnarray}
\noindent This equation differs from the corresponding equation of Ref.~\cite{kazakov2} exactly by the last term. The point is that in Ref.~\cite{kazakov2}, channel flame propagation was considered as the propagation of a $2b$-periodic infinite flame; though not stated explicitly in Ref.~\cite{kazakov2}, vanishing of $\langle M\omega_+/v^2_+\rangle$ is actually the condition of existence of improper integrals over the infinite flame front, appearing in the derivation of the rotational velocity component. The question whether or not this quantity is zero turns out to be quite nontrivial. As to its real part, it is zero identically as the average of an odd periodic function. However, there is no apparent reason why $\langle M w_+/v^2_+\rangle$ should vanish. Using the small $(\theta-1)$-expansion, it can be shown that this quantity does vanish at least to the fourth post-Sivashinsky approximation (which corresponds to keeping $O((\theta - 1)^7)$-terms in the master equation), as an average of the derivative of a periodic function \cite{kazakov5}. Furthermore, it vanishes in the opposite case of flames with large front-slope (such as flames anchored in high-velocity streams), which can be shown even without using the master equation, but we were unable to prove this in the general case. It is interesting to note that the new term, if nonzero, would mean appearance of an effective gravitational field. Indeed, in the simplest case of constant gravity parallel to the channel walls, a term proportional to $f'$ appears in the on-shell vorticity \citep{hayes}, which has the form
\begin{eqnarray}
\Delta\sigma_+ = - \frac{(\theta - 1)}{\theta N}g f'(x,t)\,,\nonumber
\end{eqnarray}
\noindent where $g$ is the gravity acceleration. Since $z'_+ = 1 + if'(x,t),$ it follows from Eq.~(\ref{masterst}) that taking roughly $u_+ \sim v_+ \sim \theta,$ the effective acceleration can be estimated as $g_{\rm eff} \simeq - \langle M w_+/v^2_+\rangle.$ In this connection, it is worth of mentioning that the extra term belongs to the kernel of the $\EuScript{H}$-operator, namely, in the general case of curved channel, $\hat{\EuScript{H}\,}g'(z_+) = 0,$ as can be easily checked using the definition (\ref{hilbert}).

\subsection*{Quasi-steady flames}

Characteristic time of spontaneous evolution of a flame disturbance with characteristic wavelength $\lambda$ is, by the order of magnitude, $T\sim \lambda/U_{\rm f},$ where $U_{\rm f}$ is the flame front velocity relative to the fresh gas. Hence, dynamics of disturbances with sufficiently large $\lambda$ is slow, and the flame can be considered quasi-steady, provided that the steady regime exists in the absence of disturbances. Specifically, since there is only one internal time scale for steady flames, namely, $\lambda_{\rm c}/U_{\rm f},$ such consideration is applicable, if $\lambda\gg \lambda_{\rm c}.$ Flames propagating in channels with slowly varying width, or in a slowly varying gravity deliver simplest examples of quasi-steady flames. The results obtained above for steady flames are readily generalized to this case.

In terms of the Fourier decomposition of the memory kernel,
$$M(\tilde{x},t) = \int_{-\infty}^{+\infty} d\omega M(\tilde{x},\omega)e^{-i\omega t},$$ slowness of flame dynamics means that $M(\tilde{x},\omega)$ is noticeable only for sufficiently small $|\omega|.$ Substituting this into Eq.~(\ref{mreg}), the $\varepsilon$-regularized expression for ${\rm Im}\,\EuScript{M}$ can be written as
\begin{eqnarray}\label{mreg1}
{\rm Im}\,\EuScript{M}(\tilde{x},t,\tau) &=& \frac{1}{2i}\lim\limits_{\varepsilon\to 0}\int_{-\infty}^{+\infty} d\omega M(\tilde{x},\omega)e^{-i\omega t}\int\limits_{\tau}^{\overline{\tau}}d\tau_1 e^{-(\varepsilon - i\omega)\tau_1} \nonumber\\ &=& \frac{1}{2i}\lim\limits_{\varepsilon\to 0}\int_{-\infty}^{+\infty} d\omega M(\tilde{x},\omega)e^{-i\omega t}\frac{e^{-(\varepsilon - i\omega)\tau} - e^{-(\varepsilon - i\omega)\overline{\tau}}}{(\varepsilon - i\omega)}\,.
\end{eqnarray}
\noindent It is seen that as a result of the flame unsteadiness, the regularizing parameter $\varepsilon$ becomes effectively complex. To sum the series in Eq.~(\ref{master}) in the steady case, we had to expand the right hand side of Eq.~(\ref{series}) in powers of $\varepsilon$ keeping terms $O(\varepsilon).$ Therefore, in order to take into account the unsteady effects to the first order in $\omega,$ we have to retain in Eq.~(\ref{series}) also the term $O(\varepsilon^2),$ and to replace $\varepsilon\to -i\omega$ afterwards. A simple calculation gives
\begin{eqnarray}&&
\left.\sum_{n=-\infty}^{+\infty}\chi(n)
\left\{e^{-\varepsilon\tau_n} - e^{-\varepsilon\overline{\tau_n}} - e^{-\varepsilon\sigma_n} + e^{-\varepsilon\overline{\sigma_n}}\right\}\right|_{\varepsilon^2} \nonumber\\&& =  \frac{\varepsilon^2}{12b}\left\{\zeta^3\left(\frac{1}{D^2} - \frac{1}{\overline{D}^2}\right) - 3\zeta^2\left(\frac{g(\tilde{z})}{D^2} + \frac{\overline{g(\tilde{z})} - 2b}{\overline{D}^2}\right) - {\rm c.c.}\right\},\nonumber
\end{eqnarray}
\noindent
where now $D = D(\tilde{x},t) = \tilde{\gamma}[w_+(\tilde{x},t) + i\hat{u}_+(\tilde{x},t)],$ and contributions linear in $\zeta$ have been omitted (as before, they fall off from Eq.~(\ref{master}) anyway). Hence,
\begin{eqnarray}&&
{\rm Im}\left.\sum_{n=-\infty}^{+\infty}\chi(n)\left\{\EuScript{M}(\tilde{x},t,\tau_n) - \EuScript{M}(\tilde{x},t,\sigma_n)\right\}\right|_{\omega} \nonumber\\ && =  \frac{1}{2i}\int_{-\infty}^{+\infty} d\omega M(\tilde{x},\omega)e^{-i\omega t}\frac{(-i\omega)}{12b}\left\{\zeta^3\left(\frac{1}{D^2} - \frac{1}{\overline{D}^2}\right) - 3\zeta^2\left(\frac{g(\tilde{z})}{D^2} + \frac{\overline{g(\tilde{z})} - 2b}{\overline{D}^2}\right) - {\rm c.c.}\right\} \nonumber\\ && =  \frac{1}{2i}\frac{\partial M(\tilde{x},t)}{\partial t}\frac{1}{12b}\left\{\dots \right\}\nonumber
\end{eqnarray}
\noindent With the same accuracy, one has $${\rm Im}\, \EuScript{M}(\tilde{x},t,\tau_0) = \frac{1}{2i}\left[M(\tilde{x},t)(\overline{\tau_0} - \tau_0) + \frac{\partial M(\tilde{x},t)}{\partial t}\frac{\tau^2_0 - \overline{\tau_0}^2}{2}\right],$$ and a similar expression for ${\rm Im}\, \EuScript{M}(\tilde{x},t,\sigma_0).$ Substituting these formulas in Eq.~(\ref{master}), and performing differentiations as before yields the master equation for quasi-steady flames
\begin{eqnarray}
2\left(\frac{\omega_-}{\gamma_+}\right)' &&\hspace{-0,3cm} + \left(1 + i\hat{\EuScript{H}\,}\right)\left\{\left(\frac{[\omega]}{\gamma_+}\right)'  - \frac{M(x,t)\hat{\omega}_+(x,t)}{\gamma_+ \hat{v}^2_+(x,t)} + \frac{g'(z_+)}{2b}\int_{-B}^{+B}d\tilde{x}\frac{M(\tilde{x},t)\hat{\omega}_+(\tilde{x},t)}{\tilde{\gamma} \hat{v}^2_+(\tilde{x},t)} \right.\nonumber\\&&\left. + \frac{ig'(z_+)}{2}\int_{-B}^{+B}d\tilde{x} \left(\frac{\hat{\omega}_+(\tilde{x},t)}{\tilde{\gamma}\hat{v}^2_+(\tilde{x},t)}\right)^2\frac{\partial M(\tilde{x},t)}{\partial t}\left[\chi(\tilde{x} - x) +\frac{g(z_+) - g(\tilde{z})}{b}\right]\right\} = 0. \nonumber
\end{eqnarray}
\noindent It is straightforward to check that the new term has zero average across the channel, as it should.

\section{The small gas expansion limit}\label{smalltheta}

\subsection{Derivation of equation for the flame front position}

Equation (\ref{master}) also considerably simplifies in the case when $(\theta-1)\equiv \alpha$ is small. Since the cutoff wavelength $\lambda_{\rm c} = O(1/\alpha),$ the wave numbers of the flame perturbations are
$O(\alpha),$ implying that spatial differentiation of a perturbed flow
variable raises its order with respect to $\alpha$ by one. Next, estimation of the perturbation growth rate using the Darrieus-Landau relation shows that it is $O(\alpha^2),$ therefore, time-differentiation adds two orders. The evolution equation  (\ref{evolutiongen}) [with $S\equiv 0$] then tells us that $u_- = O(\alpha^2),$  hence, $w_- = O(\alpha^2).$ Also, the slowness of flame dynamics means that the channel width must vary sufficiently slowly along the channel, namely, the phase of the function $\gamma(z)= dg(z)/dz$ must be small, ${\rm arg}\gamma(z)\ll 1.$ A natural representations of the maps meeting this requirement is
\begin{eqnarray}\label{natrep}
g^{-1}(\zeta) = Lh\left(\frac{\zeta}{L}\right),
\end{eqnarray}
\noindent
where $L = O(b/\alpha^2) \gg b$ is the characteristic length scale over which the channel width changes significantly, and the function $h(\zeta)$ is ``normal,'' in the sense that $|dh/d\zeta| = O(1)$ [$dh/d\zeta$ denotes hereon derivative of the function $h(\zeta),$ not $h(\zeta/L)$]. It obeys the usual reflection rule $h\left(-\overline{\zeta}\right) = - \overline{h\left(\zeta\right)}\,.$ It follows from this relation that the derivative of $g^{-1}(\zeta)$ given by Eq.~(\ref{natrep}) is a real $O(1)$-quantity at the imaginary axis ($\zeta = i\xi$), which acquires an $O(b/L)$ imaginary part at the channel walls ($\zeta = \pm b + i\xi$). Using this representation, the factor $\gamma^{-1}_+$ appearing in Eq.~(\ref{master}) can be evaluated as
\begin{eqnarray}\label{gamma}
\frac{1}{\gamma_+} = \frac{dg^{-1}}{d\zeta}(\zeta_+) = \frac{dh}{d\zeta}\left(\zeta_+/L\right) = \frac{dh}{d\zeta}\left(i\xi_0/L\right) + \frac{d^2g^{-1}}{d\zeta^2}(i\xi_0)(\zeta_+ - i\xi_0) + O\left(\frac{b^2}{L^2}\right),
\end{eqnarray}
\noindent
where $\xi_0$ is the image of the middle-point of the flame front, $i\xi_0 = g(if(0,t)).$ The quantities $$\frac{dh}{d\zeta}(i\xi_0/L)\equiv s(t), \quad \frac{1}{i}\frac{d^2g^{-1}}{d\zeta^2}(i\xi_0) \equiv k(t)$$ are thus purely real. We note for future reference that they can be related to the flame front velocity at $x=0,$ namely, differentiating $s(t),$ and using the above definitions gives
\begin{eqnarray}\label{sk}
\frac{ds(t)}{dt} = \frac{d^2 g^{-1}}{d\zeta^2}(i\xi_0)\frac{dg}{dz}(if(0,t))i\frac{d f(0,t)}{d t} = - \frac{k(t)}{s(t)}\frac{d f(0,t)}{d t} = \frac{k(t)}{s(t)} + O(\alpha^4)\,.
\end{eqnarray}
\noindent The last expression follows from the fact that the flame propagates downwards, its velocity being $1 + O(\alpha^2).$

Below, we consider the first post-Sivashinsky approximation which corresponds to retaining terms $O(\alpha^4)$ in Eq.~(\ref{master}). Taking into account that $\sigma_+ = O(\alpha^4)$ [Cf. Eq.~(\ref{vorticity})], one has to this order
$$M = \sigma_+ = - \alpha\left(\frac{\partial f'}{\partial t} + f'f''\right)\,.$$ Moreover, since $M$ is already of the fourth order, $\tau_n,\sigma_n$ can be taken in the form
$$\tau_n = is(t)[4bn - \zeta + g(\tilde{z})],\quad \sigma_n = is(t)[4bn +2b - \zeta - \overline{g(\tilde{z})}].$$
One has, within the given accuracy,
\begin{eqnarray}
\EuScript{M}(\tilde{x},t,\tau_n) - \EuScript{M}(\tilde{x},t,\sigma_n) &=& \int_{\tau_n}^{\sigma_n}d\tau_1 \sigma_+(\tilde{x},t - \tau_1) = \sigma_+(\tilde{x},t)(\sigma_n - \tau_n) \nonumber\\ &=& \sigma_+(\tilde{x},t)is(t)[2b - \overline{g(\tilde{z})} - g(\tilde{z})].\nonumber
\end{eqnarray}
\noindent This is independent of $z,$ which means that the series in Eq.~(\ref{master}) requires no care, and the image contributions can be omitted altogether. On the other hand, for the $\tau_0$-contribution one finds ${\rm Im}\,\EuScript{M}(\tilde{x},t,\tau_0) = \sigma_+(\tilde{x},t)s(t)[\zeta - g(\tilde{z})].$ It is worth of mentioning that the possibility to replace $M(\tilde{x},t-\tau)$ by $M(\tilde{x},t)$ signifies the absence of memory effects in the present case. When expanding various terms in Eq.~(\ref{master}) in powers of $\alpha,$ it is sufficient to retain only the $O(1)$-part in the expression (\ref{gamma}), the only exception being the contribution proportional to $[u].$ Namely, using Eq.~(\ref{jumps}) one has
$$\frac{[u]}{\gamma_+} = \frac{\alpha}{N}\left[s(t) + ik(t)(\zeta_+ - i\xi_0) + O(\alpha^4)\right] = \alpha s(t)\left(1 - \frac{f'^2}{2}\right) + i\alpha k(t)\eta + O(\alpha^4),$$ where we took into account that $\xi_+ - \xi_0 = O(\alpha).$ Writing also $\eta = x/s(t) + O(\alpha^2),$ substituting this into Eq.~(\ref{master}), and carrying out differentiations as in the preceding section gives
\begin{eqnarray}\label{masterps}
2\left(\omega_-\right)' + \alpha\left(1 + i\hat{\EuScript{H}\,}\right)\left\{ix\frac{k(t)}{s^2(t)} - if' + \frac{\partial f}{\partial t} \right\}' = 0.
\end{eqnarray}
\noindent Next, it follows from Eq.~(\ref{gamma}) that the value of $1/\gamma_+$ at the right channel wall is $s(t)[1 + ik(t)B/s^2(t)].$ Considered as a vector in the complex $z$-plane, this is normal to the wall, hence, $x k(t)/s^2(t) - f'$ vanishes there. Therefore, the expression in the braces in Eq.~(\ref{masterps}) has the same value for $x = \pm B.$ For such a function $a(x),$ the $\hat{\EuScript{H}\,}$-operator acting on its derivative can be rewritten as
\begin{eqnarray}
\left(\hat{\EuScript{H}\,}a'\right)(x) &=& \frac{g'(z_+)}{2b}\fint_{-B}^{+B}d\tilde{x} \cot\left\{\frac{\pi}{2b}[g(\tilde{z}) - g(z_+)]\right\}a'(\tilde{x}) \nonumber\\ &=& \frac{1}{2b}\frac{d}{dx}\fint_{-B}^{+B}d\tilde{x} g'(\tilde{z})\cot\left\{\frac{\pi}{2b}[g(\tilde{z}) - g(z_+)]\right\}a(\tilde{x}).
\nonumber
\end{eqnarray}
\noindent Furthermore, $g'(\tilde{z}) = (dg/d\tilde{z})\tilde{z}'$ can be replaced here by $\tilde{z}'/s(t),$ with an error of the order $O(\alpha^2).$ Noting also that within the same accuracy, $B = s(t)b,$ we find
\begin{eqnarray}\label{hilbert1}
\left(\hat{\EuScript{H}\,}a'\right)(x) = \frac{1}{2B}\frac{d}{dx}\fint_{-B}^{+B}d\tilde{z} \cot\left\{\frac{\pi}{2B}[\tilde{z} - z_+]\right\}a(\tilde{x}).
\end{eqnarray}
\noindent But the latter expression can be shown to be just $(\hat{H}_Ba)',$ again up to terms of the second relative order in $\alpha,$ where $\hat{H}_B$ is the usual Hilbert operator defined on the interval $(-B,+B)$ (Cf.~Eq.~(B12) in Ref.~\cite{kazakov2})
\begin{eqnarray}\label{hilbertb}
\left(\hat{H}_B a\right)(x) = \frac{1}{2B}\fint_{-B}^{+B}d\tilde{x} \cot\left\{\frac{\pi}{2B}[\tilde{x} - x]\right\}a(\tilde{x}).
\end{eqnarray}
\noindent Integration of Eq.~(\ref{masterps}) thus yields
\begin{eqnarray}
2\omega_- + \alpha\left(1 + i\hat{H}_B\right)\left\{ix\frac{k(t)}{s^2(t)} - if' + \frac{\partial f}{\partial t}\right\} = C(t),\nonumber
\end{eqnarray}
\noindent where $C(t)$ is an ``integration constant.'' The parity properties (\ref{reflect1}) imply that it is real. Taking into account that the Hilbert operator is also real, extraction of the real and imaginary parts of this equation gives
\begin{eqnarray}\label{masterpsu}
2u_- + \alpha\hat{H}_B\left\{f' - x\frac{k(t)}{s^2(t)}\right\} + \alpha \frac{\partial f}{\partial t} = C(t), \\
2w_-  - \alpha f' = O(\alpha^3). \label{masterw}
\end{eqnarray}
\noindent The latter equation is written out up to $O(\alpha^3)$-terms, because the $w$-component of the fresh gas velocity is multiplied by $f'=O(\alpha)$ in the evolution equation (\ref{evolutiongen}) which in the present case reads
\begin{eqnarray}\label{evolution}
u_- - f' w_- = \frac{\partial f}{\partial t} + 1 + \frac{f'^2}{2}\,.
\end{eqnarray}
\noindent
Combining the last three equations, we arrive at the following equation for the flame front position
\begin{eqnarray}\label{fposition}
(\theta +1)\frac{\partial f}{\partial t} + \theta f'^2 = -(\theta - 1)\hat{H}_B\left\{f' - x\frac{k(t)}{s^2(t)}\right\} + C(t) - 2.
\end{eqnarray}
\noindent It is convenient to introduce a new function
$$F(x,t) = f(x,t) - \frac{k(t)x^2}{2s^2(t)}\,, \quad F'(\pm B,t)=0,$$ in terms of which Eq.~(\ref{fposition}) takes the form, within the third order accuracy,
\begin{eqnarray}\label{fposition1}
(\theta + 1)\frac{\partial F}{\partial t} + \theta F'^2 + \frac{2 k(t)}{s^2(t)}xF' = - (\theta - 1)\hat{H}_B F' + C(t) - 2.
\end{eqnarray}
\noindent Averaging this equation along the front yields the function $C(t)$ itself $$C(t) = 2 + (\theta +1)\frac{\partial \langle F\rangle}{\partial t} + \theta \langle F'^2\rangle + \frac{2 k(t)}{s^2(t)}\langle xF'\rangle.$$ In the case of a straight channel, $k(t) \equiv 0,$ and the obtained equation is simply the Sivashinsky-Clavin equation \citep{sivclav} (with the term $C(t)$ added according to \cite{joulin1}). Equation (\ref{fposition1}) is valid for zero-thickness flames. Account of the transport processes inside the flame front adds higher-derivative terms proportional to the cutoff $\lambda_{\rm c}.$ If, as usual, these processes can be considered linear, then the term to be added to the right hand side of Eq.~(\ref{fposition}) has the form \cite{siv1,pelce} $(\theta - 1)\lambda_{\rm c} f''/2\pi,$ and hence, Eq.~(\ref{fposition1}) is to be replaced by
\begin{eqnarray}\label{fposition2}
(\theta + 1)\left[\frac{\partial F}{\partial t} + \frac{ds}{dt}\frac{xF'}{s(t)}\right] + \theta F'^2 = - (\theta - 1)\hat{H}_B F'  + (\theta - 1)\frac{\lambda_{\rm c}}{2\pi}F'' + \tilde{C}(t)\,,
\end{eqnarray}
\noindent where
\begin{eqnarray}
\tilde{C}(t) \equiv C(t) - 2 + (\theta - 1)\frac{\lambda_{\rm c}}{2\pi}\frac{k(t)}{s^2(t)}\,,\nonumber
\end{eqnarray}
\noindent
and we also used Eq.~(\ref{sk}) to rearrange the left hand side.

Concerning the structure of the obtained equation, it is important to note that although the function $s(t)$ is related by its definition to the front location within the channel, it is actually independent of the specifics of flame evolution. Namely, it can be written as
\begin{eqnarray}\label{sfunction}
s(t) = \frac{dh}{d\zeta}\left(\frac{g(if_0 - it)}{L}\right),
\end{eqnarray}
\noindent where $f_0 \equiv f(0,0).$ Since the flame velocity increase due to the front curvature is $O(\alpha^2),$ corrections to this formula are of the fourth order. Thus, in the first post-Sivashinsky approximation, the function $s(t)$ is to be considered as non-dynamical quantity predetermined by the channel geometry and  initial front position.

\subsection{Front evolution in terms of pole dynamics}

Though not obvious from its present form, Eq.~(\ref{fposition2}) is amenable to the pole decomposition. To see this, let us introduce new variables $\eta,\Phi$ according to
$$F(x,t) = \Phi\left(\frac{x}{s(t)},t\right), \quad \eta = \frac{x}{s(t)}\,.$$ Then the expression in the square brackets on the left of Eq.~(\ref{fposition2}) is just $\partial\Phi/\partial t,$ so that this equation becomes
\begin{eqnarray}\label{fpos3}
(\theta + 1)\frac{\partial \Phi}{\partial t} + \frac{\theta}{s^2(t)} \left(\frac{\partial\Phi}{\partial\eta}\right)^2 = - \frac{\theta - 1}{s(t)}\hat{H}_b \frac{\partial\Phi}{\partial\eta}  + \frac{\theta - 1}{s^2(t)}\frac{\lambda_{\rm c}}{2\pi}\frac{\partial^2\Phi}{\partial\eta^2} + \tilde{C}(t)\,,
\end{eqnarray}
\noindent where $\hat{H}_b$ is defined by
\begin{eqnarray}
\left(\hat{H}_b a\right)(\eta) = \frac{1}{2b}\fint_{-b}^{+b}d\tilde{\eta} \cot\left\{\frac{\pi}{2b}[\tilde{\eta} - \eta]\right\}a(\tilde{\eta}).\nonumber
\end{eqnarray}
\noindent
Solutions of Eq.~(\ref{fpos3}) can be found in the form \cite{thual,joulin1991}
\begin{eqnarray}\label{anzats}
\Phi(\eta,t) = \Phi_0(t) + A \sum_{k = 1}^{2 P}
\ln\sin\left\{\frac{\pi}{2 b}[\eta - \eta_k(t)]\right\}.
\end{eqnarray}
\noindent Here $P$ is the number of complex-conjugate pairs of poles located at the points $\eta_k,$ $k = 1,...,2P,$ in the complex $\eta$-plane. To determine the amplitude $A$ and the functions $\Phi_0(t),$ $\eta_k(t),$ we substitute this anzats into Eq.~(\ref{fpos3}), and use the formulas
\begin{eqnarray}
\hat{H}_b\frac{\partial\Phi}{\partial\eta} &=& - \frac{\pi A}{2 b}\sum_{k = 1}^{2 P}\left(1 + i\chi({\rm Im}\,\eta_k)\cot\left\{\frac{\pi}{2 b}[\eta - \eta_k(t)]\right\}\right), \nonumber\\ \cot x \cot y &=& -1 + \cot(x - y
)(\cot y - \cot x)\,.\label{cots}
\end{eqnarray}
\noindent This leads to the following system of ordinary differential equations
\begin{eqnarray}\label{solution1}
A &=& - \frac{\theta - 1 }{\theta}\frac{\lambda_{\rm c}}{2\pi}\,,
\\ \frac{d\Phi_0}{dt} &=& \frac{(\theta - 1)^2}{\theta(\theta + 1)}\frac{P\lambda_{\rm c}}{2 b s(t)}\left(\frac{P\lambda_{\rm c}}{2 b s(t)} - 1\right) + \frac{\tilde{C}(t)}{(\theta + 1)},\label{solution2}\\
\frac{d\eta_k}{dt} &=& - \frac{\theta - 1}{\theta + 1}\frac{1}{s(t)}\left(i\chi({\rm Im}\,\eta_k) + \frac{\lambda_{\rm c}}{2
b s(t)}\sum\limits_{\genfrac{}{}{0pt}{}{m = 1}{m\ne k}}^{2
P}\cot\left\{\frac{\pi}{2 b}(\eta_k - \eta_m)\right\}\right). \label{solution3}
\end{eqnarray}
\noindent As is seen, this system does not determine the functions $\Phi_0(t),$ $\tilde{C}(t)$ individually, but only a linear combination thereof. This corresponds to the fact already mentioned above that the master equation has one and the same form whatever velocity of the incoming fresh gas (potential) flow. We now recall that the fresh gas is assumed to be initially at rest. The flow continuity then implies that its total flux through the front is zero. Using the evolution equation (\ref{evolution}), this condition can be written as
\begin{eqnarray}\label{fix}
\left\langle \frac{\partial f}{\partial t}\right\rangle = - 1 - \frac{\left\langle f'^2\right\rangle}{2}\,.
\end{eqnarray}
\noindent In terms of the original variables, Eq.~(\ref{anzats}) reads
\begin{eqnarray}\label{anzats1}
f(x,t) = \Phi_0(t) + \frac{x^2}{2s(t)}\frac{ds}{dt} + A \sum_{k = 1}^{2 P}
\ln\sin\left\{\frac{\pi}{2 b}\left[\frac{x}{s(t)} - \eta_k(t)\right]\right\}.
\end{eqnarray}
\noindent Substituting this into Eq.~(\ref{fix}) gives, within the given approximation,
\begin{eqnarray}&&
\frac{d\Phi_0}{dt} - \sum_{k = 1}^{2 P}\frac{\pi A}{2b}\frac{d\eta_k}{dt}\frac{1}{2B}\int_{-B}^{+B}d\tilde{x}\cot\left\{\frac{\pi}{2 b}\left[\frac{\tilde{x}}{s} - \eta_k\right]\right\}\nonumber\\&& = -1 - \frac{1}{2}\left(\frac{\pi A}{2bs}\right)^2\frac{1}{2B}\int_{-B}^{+B}d\tilde{x}\left(\sum_{k = 1}^{2 P}\cot\left\{\frac{\pi}{2 b}\left[\frac{\tilde{x}}{s} - \eta_k\right]\right\}\right)^2 \nonumber
\end{eqnarray}
\noindent (notice cancelation of terms proportional to $ds/dt$). A straightforward calculation using the formulas (\ref{cots}), (\ref{solution3}), and
$$\left\langle \cot\left\{\frac{\pi}{2 b}\left[\frac{x}{s} - \eta_k\right]\right\}\right\rangle = i\chi({\rm Im}\,\eta_k)$$
gives
$$\frac{d\Phi_0}{dt} = - 1 - \frac{(\theta - 1)^2}{2\theta^2}\frac{P\lambda_{\rm c}}{2 b s(t)}\left(1 - \frac{P\lambda_{\rm c}}{2 b s(t)}\right) + \frac{(\theta - 1)^2}{2\theta^2}\frac{\lambda_{\rm c}}{2b}\sum_{k = 1}^{P}\frac{d\,{\rm Im}\,\eta_k}{dt}\,,$$ where the latter sum runs only over the poles in the upper half-plane. Integrating this relation with the same accuracy, we find
$$\Phi_0(t) = - t - t\frac{(\theta - 1)^2}{2\theta^2}\frac{P\lambda_{\rm c}}{2 b s(t)}\left(1 - \frac{P\lambda_{\rm c}}{2 b s(t)}\right) + \frac{(\theta - 1)^2}{2\theta^2}\frac{\lambda_{\rm c}}{2b}\sum_{k = 1}^{P}{\rm Im}\,\eta_k(t) + {\rm const},$$
where the additive constant determines initial front position.

To summarize, in order to determine the flame front evolution in a channel, one uses the given map $g^{-1}(\zeta)$ and the quantity $f_0$ to construct the function $s(t)$ according to Eq.~(\ref{sfunction}), and then solves the ordinary differential equations (\ref{solution3}) for $\eta_k(t).$ Substituting the latter into Eq.~(\ref{anzats1}) yields the function $f(x,t).$ An important point of this procedure is the proper choice of the number $P.$ Since by the construction the number of pole pairs is independent of time, any choice of $P$ predetermines to a certain extent the flame evolution. At the same time, it is known that in straight channels, there is only one stable steady pole solution for every channel width \cite{matalon1,matalon2}. The corresponding number of pole pairs is given by
$$P_{\rm s} = {\rm Int} \left(\frac{B}{\lambda_{\rm c}} + \frac{1}{2}\right),$$
where ${\rm Int}(x)$ denotes the integer part of $x.$ Thus, in order to construct the true solution which correctly describes unstable flame behavior in curved channels, $P$ is to be taken large enough. In channels of finite width, it is sufficient to set $P_{\rm s} = {\rm Int} \left(B_{{\rm max}}/\lambda_{\rm c} + 1/2\right),$ where $2B_{{\rm max}}$ is the maximal channel width. Then in regions with $B < B_{{\rm max}},$ a number of redundant poles move to infinity in the $\eta$-plane. In diverging channels, however, $P_{\rm s}$ is formally infinite, so that this case requires special consideration.

\section{Discussion and conclusions}\label{conclude}

The results obtained in this paper extend the on-shell flame description constructed in Refs.~\cite{kazakov1,kazakov2,jerk1,jerk2} to the case of flames propagating in curved symmetric channels. Namely, complex Eq.~(\ref{master}) relates the on-shell gas velocity and front position of thin unsteady flame in the most general form. This equation shows that the influence of channel geometry on flame dynamics is deeper and more diverse than might be expected from simple considerations based on potential flow models \cite{siv1,frankel1990}. In the latter, the effect of wall curvature is  contained entirely in the quantity $1/\gamma_+$ accompanying $\omega_-$ and $[\omega]$ in Eq.~(\ref{master}), which is just a conversion factor relating on-shell velocities of the potential flows in the physical channel and its image. The master equation shows that things are actually much more complicated. In addition to the usual nonlocality related to potential flows and described by the operator $\hat{\EuScript{H}\,},$ vorticity produced by the curved flame brings in specific nonlocality associated with the integral term in Eq.~(\ref{master}). The structure of this term essentially depends on the properties of the map $g(z).$ Moreover, temporal nonlocality described by the memory kernel is also affected by these properties through the $\tau$- and $\sigma$-poles appearing in the function $\EuScript{M}(\tilde{x},t,\tau).$

It is worth of recalling that the derivation of Eq.~(\ref{master}) relies on two important technical assumptions, namely, convergence of the integrals over $\tau$ corresponding to the remote past, and convergence of the series over the $\tau$- and $\sigma$-poles in Eq.~(\ref{vint6}). Concerning the former, asymptotic straightness of the channel is a sufficient condition, as was discussed in Sec.~\ref{integrep}, but more general channel configurations may require additional consideration such as for instance stability analysis of a particular jet configuration. As to the series convergence, it seems to pose no serious problem as it can be naturally ensured on physical grounds. Indeed, as was demonstrated in Sec.~\ref{steady}, even in the case of a steady flame when assumptions about the large-time behavior of the memory kernel are irrelevant, convergence is easily achieved taking into account small gas viscosity.

Finally, we mention that flame anchoring systems such as metallic rods can be easily incorporated into the above construction following the lines of Ref.~\cite{jerk3}. In particular, this gives immediate analytic access to the problems involving fast flow burning, because the $\EuScript{H}$-operator considerably simplifies in this case (see Refs.~\cite{jerk3,kazakov4} for details).

\acknowledgments{The work presented in this paper was carried out
at the {\it Laboratoire de Combustion et de D\'etonique}. One of the
authors (K.A.K.) thanks the {\it Centre National de la Recherche
Scientifique} for supporting his stay at the Laboratory as a {\it
Chercheur Associ\'e}.}

\begin{appendix}

\section*{Appendix. Proof of the identity $\hat{\EuScript{H}\,}^2 = -1$}

\begin{figure}
\centering
\includegraphics[width=0.6\textwidth]{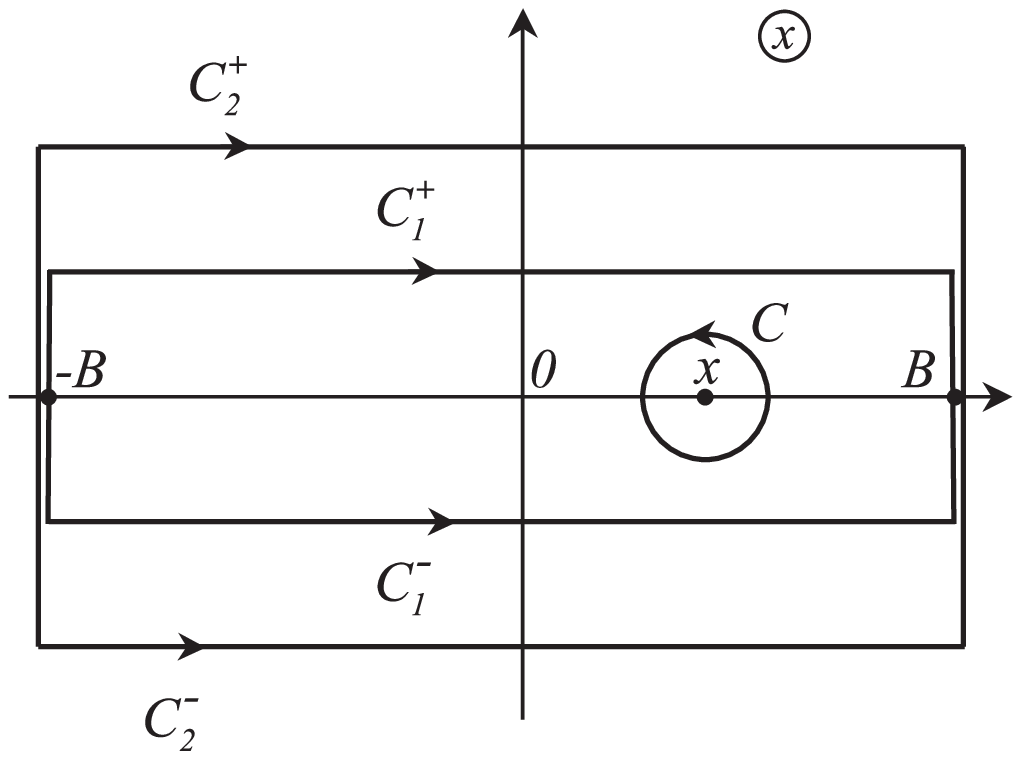}
\caption{}\label{fig4}
\end{figure}

Let us prove that $\hat{\EuScript{H}\,}$ satisfies the operator identity
$\hat{\EuScript{H}\,}^2 = - 1$ valid on the space of functions $a(x)$ which are analytic in $x$ in a vicinity of the real axis, and have zero average across the channel (the proof follows closely that given in Ref.~\cite{kazakov2} in the case of straight channel). For this purpose we rewrite its definition as
\begin{eqnarray}\label{hilberta1}
\left(\hat{\EuScript{H}\,}a\right)(x) = \frac{g'(z_+)}{4b}\int_{C_1}d x_1 \cot\left\{\frac{\pi}{2b}[g(z_1) - g(z_+)]\right\}a(x_1), \quad z_1 = x_1 + if(x_1,t),
\end{eqnarray}
\noindent where $C_1 = C_1^- \cup
C_1^+$ is the contour in the complex $\tilde{x}$-plane, shown on
Fig.~\ref{fig4}. We then have
\begin{eqnarray}&&
\left(\hat{\EuScript{H}\,}^2 a\right)(x) = \left(\hat{\EuScript{H}\,} (\hat{\EuScript{H}\,}a)\right)(x) \nonumber\\&& = \frac{g'(z_+)}{4b}\int\limits_{C_1}dx_1\cot\left\{\frac{\pi}{2b}[g(z_1) - g(z_+)]\right\}\frac{g'(z_1)}{4b}
\int\limits_{C_2}dx_2\cot\left\{\frac{\pi}{2b}[g(z_2) - g(z_1)]\right\}a(x_2)\,,\nonumber
\end{eqnarray}
\noindent where $z_2 = x_2 + if(x_2,t),$ and the contour $C_2 = C_2^- \cup C_2^+$ of
integration over $x_2$ comprises $C_1$ (see Fig.~\ref{fig4}). The contours $C_1,$ $C_2$ are chosen so that all singularities of the integrand (if any), except the poles of the cotangent, remain above $C^+_{2},$ or below $C^-_{2},$ which is possible under our assumptions about analytic properties of $a(x),$ $g(z).$
Now we change the order of integration, and use the formula (\ref{cots}) to compute the integral over $x_1$
\begin{eqnarray}&&
\int\limits_{C_1}\frac{dg(z_1)}{4b}\cot\left\{\frac{\pi}{2b}[g(z_1) - g(z_+)]\right\}\cot\left\{\frac{\pi}{2b}[g(z_2) - g(z_1)]\right\} \nonumber\\&&
= 1 - \frac{1}{2\pi}\cot\left\{\frac{\pi}{2b}[g(z_2) - g(z_+)]\right\}\int\limits_{C_1}d\ln \frac{\sin \left\{\displaystyle\frac{\pi}{2b}[g(z_1) - g(z_2)]\right\}}{\sin \left\{\displaystyle\frac{\pi}{2b}[g(z_1) - g(z_+)]\right\}}\,. \nonumber
\end{eqnarray}
\noindent In view of the $2b$-periodicity of the integrand, it is phase change of the argument of the logarithm that only contributes. It is equal to $2\pi i$ when $x_1$ runs $C^+_1$ and $x_2\in C^+_2$; $(-2\pi i)$ when $x_1$ runs $C^-_1$ and $x_2\in C^-_2,$ and zero in the other two cases. Taking into account also Eq.~(\ref{average}), we thus find
\begin{eqnarray}\label{apb4}&&
\left(\hat{\EuScript{H}\,}^2 a\right)(x) = \frac{ig'(z_+)}{4b}\left[\int\limits_{C_2^-} - \int\limits_{C_2^+}\right]dx_2a(x_2)\cot\left\{\frac{\pi}{2b}[g(z_2) - g(z_+)]\right\} \nonumber\\&& = \frac{ig'(z_+)}{4b}\int\limits_C d
x_2a(x_2)\cot\left\{\frac{\pi}{2b}[g(z_2) - g(z_+)]\right\} \nonumber\\&& = \frac{ig'(z_+)}{4b}2\pi i\cdot{\rm
res}\left.a(x_2)\cot\left\{\frac{\pi}{2b}[g(z_2) - g(z_+)]\right\}\right|_{x_2 = x} = - a(x)\,,
\end{eqnarray}
\noindent as was to be proved.

\end{appendix}

\bibliography{references}
~\newpage

~\\\\
\centerline{\large\bf List of figures}\\\\  

\hspace{-1,5cm} Fig.1: Curved channel in the physical $z$-plane, and its image in the auxiliary $\zeta$-plane. \\ Flame propagates downwards.
\dotfill 24\\

\hspace{-1,5cm} Fig.2: Contour of integration in the complex $\tau$-plane in Eq.~(\ref{vint5}), and general location of $\tau$- and $\sigma$-poles. \\ Indented is the region near the pole $\tau_{-n}$ where the argument of the cotangent has imaginary part $\sim b.$
\dotfill 25\\ 

\hspace{-1,5cm} Fig.3: Flame stabilized by incoming fresh gas flow in a bottle-shaped channel \\  corresponding to $B_{\infty} = 3,$ $b=1$ in Eq.~(\ref{bottle}).
\dotfill 26\\  

\hspace{-1,5cm} Fig.4: Contours of integration in Eqs.~(\ref{hilberta1}), (\ref{apb4}). \dotfill 27\\

~\newpage
~\newpage
~\newpage
~\newpage

\end{document}